\documentclass[journal,one column]{IEEEtran}
\usepackage{amsmath, amsthm, amsfonts}
\usepackage[english]{babel}
\usepackage[dvips]{graphicx}
\usepackage{graphicx}
\usepackage{tabularx}
\usepackage{enumerate}
\usepackage{mathtools}
\usepackage[ruled]{algorithm2e}
\usepackage{algorithmic}
\usepackage{tikz}
\usepackage{caption}
\usepackage{subfigure}
\usepackage{multirow}
\usepackage{adjustbox}
\usepackage{tablefootnote}
\usepackage{threeparttable}
\usepackage{amssymb}
\usepackage{makecell}
\addto\captionsenglish{\renewcommand{\figurename}{Fig.}}

\newcommand{\pgftextcircled}[1]{
    \setbox0=\hbox{#1}%
    \dimen0\wd0%
    \divide\dimen0 by 2%
    \begin{tikzpicture}[baseline=(a.base)]%
        \useasboundingbox (-\the\dimen0,0pt) rectangle (\the\dimen0,1pt);
        \node[circle,draw,outer sep=0pt,inner sep=0.1ex] (a) {#1};
    \end{tikzpicture}
}
\begin{document}

\title{Algorithm/Architecture Co-design of Proportionate-type LMS Adaptive Filters for Sparse System Identification}

\author{Subrahmanyam~Mula,~Vinay~Chakravarthi~Gogineni,~Anindya~Sundar~Dhar~\IEEEmembership{Member,~IEEE,}
      \thanks{The authors are with the with the Department of Electronics and Electrical Communication Engineering, Indian Institute of Technology (IIT) Kharagpur, West Bengal 721302, India, Email: svmula@iitkgp.ac.in.}}


\maketitle
\thispagestyle{empty}

\begin{abstract}
This paper investigates the problem of implementing
proportionate-type LMS family of algorithms in hardware for sparse
adaptive filtering applications especially the network echo
cancelation. We derive a re-formulated proportionate type algorithm
through algorithm-architecture co-design methodology that can be
pipelined and has an efficient architecture for hardware
implementation. We study the convergence, steady state and tracking
performances of these re-formulated algorithms for white, color and
speech inputs before implementing them in hardware. To the best of
our knowledge this is the first attempt to implement
proportionate-type  algorithms in hardware.  We show that Delayed
$\mu$-law Proportionate LMS (DMPLMS) algorithm for white input and
Delayed Wavelet MPLMS (DWMPLMS) for colored input are the robust
VLSI solutions for network echo cancelation where the sparsity of
the echo paths can vary with time. We implemented all the designs
considering $16$-bit fixed point representation in hardware,
synthesized the designs and synthesis results show that DMPLMS
algorithm with $\approx25\%$ increase in hardware over conventional
DLMS architecture, achieves $3X$ improvement in convergence rate for
white input and DWMPLMS algorithm with $\approx58\%$ increase in
hardware achieves $15X$ improvement in convergence rate for
correlated input conditions.

\end{abstract}


\section{Introduction}
Many real-life systems such as network echo cancelation $\cite{Wein}$, underwater
communication $\cite{Stoj}$ and HDTV terrestrial transmission
$\cite{Fan}$ exhibit impulse responses which are often sparse and sparse system identification $\cite{1}$
became an important research area in the last decade with the invention of proportionate-type algorithms.
First member of Proportionate-type LMS family of algorithms is the Proportionate Normalized Least Mean Square (PNLMS) algorithm $\cite{dutt2}$, which updates the filter coefficients by assigning a gain proportional to the magnitude of the current coefficient. The PNLMS algorithm has been shown to outperform the LMS and NLMS $\cite{haykin}$ algorithms when operating on a sparse impulse response.

However, PNLMS algorithm performance degrades and becomes worse than the NLMS
algorithm when the impulse response is dispersive. Several improved
PNLMS algorithms $\cite{gay}$- $\cite{deng_ipn}$ were proposed in
literature to address this issue and to make the algorithms more
robust against varying sparsity.


Another set of algorithms was designed by seeking a condition to
achieve the fastest overall convergence when all coefficients reach
the $\epsilon$-vicinity of their true values simultaneously (where
$\epsilon$ is a small positive number). This approach results in the
$\mu$-law PNLMS (MPNLMS) $\cite{mplms}$ and its variant
$\epsilon$-law PNLMS $\cite{eplms}$ (EPNLMS). The MPNLMS algorithm
addresses the issue of assigning too much update gain to large
coefficients, which occurs in the PNLMS algorithms. Even the MPNLMS
convergence rate becomes prohibitively slow for correlated input
conditions such as speech. Wavelet MPNLMS (WMPNLMS)
$\cite{deng_wave}$ is designed to address this issue by
de-correlating the input at the same time preserving the sparseness
of the impulse response.

As we can see significant research effort has been dedicated to the
development of high-performance adaptive algorithms based on
proportionate adaptation. However, much less is known about their
optimized implementation in dedicated hardware because of the huge
computational penalty. In this paper we try to address this gap. To the best of our knowledge this is the first
attempt to implement proportionate type algorithms in hardware. we
make several reformulations to the original PNLMS algorithm to make
it VLSI friendly and the reformulated algorithm is implemented in
hardware. Our main contributions include :
\begin{enumerate}
\item Various proportionate-type LMS algorithms and their VLSI implementation aspects are studied in detail.
\item Proposed DMPLMS for white input and DWMPLMS algorithm for color input through algorithm-architecture co-design and we show that the entailed loss in performance due to complexity reduction is negligible.
\item  Proposed a novel multiplier-less $3$-level sliding HAAR wavelet transform for DWMPLMS algorithm, to exploit the redundancies that exist in  wavelet computation of streaming input samples .
\item  As a proof-of-concept, we provide a synthesis study in $180$nm-CMOS technology of the proposed DPLMS, DMPLMS and DWMPLMS architectures and compare the improvement in rate of convergence vs hardware efficiency.
\end{enumerate}

 The rest of the paper is organized as follows. In the next section,
we explain the motivation and formulate the problem. In Section III,
we propose the algorithmic re-formulations and the corresponding VLSI architecture.
Section IV deals with the transform domain proportionate type
algorithms and their VLSI implementation aspects. Simulation results are presented in Section V and ASIC synthesis results are
presented in Section VI and we conclude the paper in Section VII.

\section{Background and Motivation}
\subsection{Review of Proportionate-type NLMS (Pt-NLMS) Algorithms}
Consider the problem of identifying an unknown sparse system modeled by the $L$ tap coefficient vector
$\textbf{w}_{opt}$ which takes a signal $u(n)$ with variance
$\sigma_{u}^{2}$ as input and produces the observable output
$d(n)=\textbf{w}_{opt}^{T}\textbf{u}(n)+v(n)$, where
$\textbf{u}(n)=[ u(n), u(n-1), . . . , u(n-L+1)]^{T}$ is the input
regressor and $v(n)$ is an observation noise with variance
$\sigma_{v}^{2}$ which is assumed to be white and independent of
$u(m)$, for all m, n. Then the Pt-NLMS algorithm iteratively updates the filter coefficient vector,
$\textbf{w}=[w_{0}, w_{1}, . . .,w_{L-1}]^{T}$ as follows:
\begin{equation}
\textbf{w}(n+1)=\textbf{w}(n)+\mu \frac{\textbf{G}(n)\textbf{u}(n)e(n)}{\textbf{u}^{T}(n)\textbf{G}(n)\textbf{u}(n)+\delta_{p}},
\end{equation}
where the estimated error $e(n) = d(n)-y(n)$ with filter output $y(n) =
\textbf{w}^{T}(n) \textbf{u}(n)$ and $\mu$ is the adaptation step
size. $\delta_{p}$ is the regularization parameter, and
$\textbf{G}(n)$ is the gain matrix. This gain matrix is the key
factor here which distinguishes Pt-NLMS from conventional NLMS. The gain matrix $\textbf{G}(n)$ is diagonal i.e., $\textbf{G}(n)=diag (g_{0}(n),g_{1}(n),...g_{L-1}(n))$, with $g_{i}(n) \propto |w_{i}(n)|$ is the gain factor for the tap $i$. For Proportionate NLMS, $g_{i}(n)$ is evaluated as follows:
\begin{equation}
g_{i}(n)=\frac{\gamma_{i}(n)}{\frac{1}{L} \sum\limits_{i=0}^{L-1}\gamma_{i}(n)}, \mspace{8mu} 0\leq i \leq (L-1),
\end{equation}
where
\begin{equation}
\gamma_{i}(n)=max[\rho\mspace{4mu} \gamma_{min}(n), \mathrm{F}[|w_{i}(n)|]]
\end{equation}
\begin{equation}
\gamma_{min}(n)=max(\delta, \mathrm{F}[|w_{0}(n)|],...,\mathrm{F}[|w_{L-1}(n)|],
\end{equation}
 $\mathrm{F}[|w_{i}(n)|] = |w_{i}(n)|$ for PNLMS. here $\delta$ is used to prevent the coefficients from stalling
during initialization stage when all the coefficients are reset to
$0$. The parameter $\rho$ ensures minimum gain to inactive coefficients.
It can be seen that if the current magnitude of a coefficient is large, a large step size
parameter will be assigned, where as for a small coefficient the
proportionate step size is small. In this way it emphasizes the
large coefficients to speed up their convergence, so it demonstrates
very fast initial convergence for sparse impulse response. However, this performance improvement comes at the cost of complexity.
In their $2004$ Freescale application note $\cite{freescale}$ Dybe \emph{et.al.} felt that because of a significant computational penalty imposed by the modified adaptation formula combined with the generation of matrix $G(n)$, the PNLMS algorithm appears to be most suitable for implementation in an ASIC technology. However, Jie, Chen in his PhD dissertation in $2008$ $\cite{chen_phd}$ stated that 'The high complexity associated with both PNLMS and IPNLMS algorithms makes them unsuitable for hardware implementation'. In the light of these two observations, we carry out a detailed complexity analysis of the PNLMS algorithm to show its inapplicability for VLSI implementation in its native form.
%
%



\subsection{Complexity Analysis}

There are four main tasks performed in each iteration of the PNLMS which are elaborated below and
all these steps need to be completed for one iteration before proceeding
to the next iteration.

\begin{enumerate}
\item Filter output calculation: This step is basically computing the inner product of input vector with the weight vector which requires L multiplications which can happen in parallel and adding all the partial results in an adder tree structure which require L adders and has a time complexity of $T_{mult} + \log_{2}(L)T_{add}$.

\item Weighted euclidian norm calculation : This step is to calculate the vector matrix vector product
$\textbf{u}^{T}(n)\textbf{G}(n)\textbf{u}(n)+\delta_{p}$ in each iteration which requires $2$L multiplications, L additions and has time complexity of $2T_{mult} + (1+log_{2}(L))T_{add}$

\item Gain Calculation : This has two max finding steps, one multiplication, one division, one $F(.)$ evaluation and summation of L terms and has time complexity of $2log_{2}(L) T_{cmp} + T_{mult}+T_{div}+T_{F_{eval}}$.

\item Weight update : This requires L multiplications and L additions and all of them can happen in parallel. Thus time complexity is $T_{mult} + T_{add}$.
\end{enumerate}

The complexities of all the steps are summarized in Table 1 and
Table 2. We see that time complexities of all the steps add up for
the critical path which is shown below:
\begin{align}
T_{crit} = 5T_{mult} + (2+3log_{2}(L))T_{add} + 2log_{2}(L)T_{cmp} +
T_{div} + T_{F\_eval} \nonumber.
\end{align}
This severely limits the applicability of the original PNLMS
algorithm for real-time VLSI implementations which have stringent
throughput requirements. In order to achieve an efficient VLSI
implementation, actions must be taken at the algorithm level to
further simplify the complexity characteristic of PNLMS (without compromising the
performance) before going into the architecture design. We discuss these simplifications in
the next section.

\begin{table}[h!]
\caption{Area Complexity of PNLMS}
\centering
\label{table:1}
\begin{adjustbox}{max width=\textwidth}
\begin{tabular}{lcccr}
\Xhline{2\arrayrulewidth}
  Step & Mult & Div & Add & Cmp\\
\Xhline{2\arrayrulewidth}
 Filter output & L & 0 & L & 0  \\
  \hline
 Weighted normalization & 2L & 0 & L & 0  \\
 \hline
 Weight update & 2L & 1 & L & 0  \\
 \hline
 Gain calculation & 2 & 1 & L & 2L \\
\Xhline{2\arrayrulewidth}
\end{tabular}
\end{adjustbox}
\end{table}

\begin{table}[h!]
\caption{Time Complexity of PNLMS }
\centering
\label{table:1}
\begin{tabular}{ lr }
 \Xhline{2\arrayrulewidth}
 Step & Critical path  \\
 \Xhline{2\arrayrulewidth}
 Filter output & $T_{mult} + \log_{2}(L)$ \\
  \hline
 Weighted normalization & $2T_{mult} + (1+log_{2}(L))T_{add}$  \\
 \hline
 Weight update & $T_{mult} + T_{add}$   \\
 \hline
 Gain calculation & $2log_{2}(L) T_{cmp} + T_{mult}+T_{div}+T_{F_{eval}}$ \\
 \Xhline{2\arrayrulewidth}
\end{tabular}

\end{table}

%
%
%
%

\section{Algorithmic Re-formulations and proposed architecture}
In this section we make several reformulations to the original PNLMS
algorithm to make it VLSI friendly without compromising the
performance of the algorithm. To ensure this, we compare the performance of the algorithm after each simplification with that of the original algorithm and finally we design a low complexity VLSI architecture for the re-formulated algorithm.
\subsection{Simplified Gain Calculation}
The first simplification is in the calculation of the gain calculation. For the simplified PNLMS algorithm, the gain factors are evaluated as follows:
\begin{equation}
\label{gain_eq}
g_{i}(n)=\frac{\gamma_{i}(n)}{\sum\limits_{i=0}^{L-1}\gamma_{i}(n)},
\mspace{8mu} 0\leq i \leq (L-1),
\end{equation}
where
\begin{equation}
\gamma_{i}(n)=
\mathrm{F}[(w_{i}(n) +\rho)],
\end{equation}
and
\begin{equation}
\mathrm{F}[(w_{i}(n) = |w_{i}(n)|.
\end{equation}
the parameter $\rho$ is a small positive constant added to avoid the
stalling when all the tap weights are zero at the reset and also
to ensure minimum gain to the inactive coefficients. We can see that
even with this simplified gain calculation, the adaptation gain is
proportional to the magnitude of the filter tap at time index $n$.
The simplified gain calculation avoids the usage of the $max$ functions
which are employed in the original PNLMS, there by reduces the time
complexity. We can use an adder tree structure for adding the absolute weights of all the taps to get the denominator of Eq.
\ref{gain_eq} and then the reciprocal of this sum can be fed to all the taps for proportional gain calculation.

\subsection{Proportionate LMS}
\begin{figure}[h!]
\centering
\includegraphics [height=70mm,width=75mm]{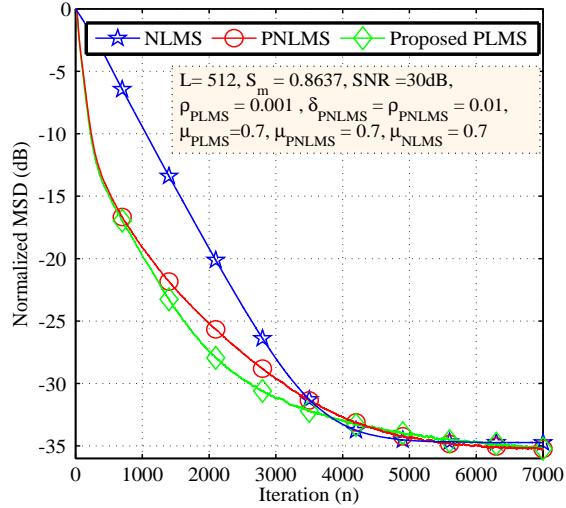}
\caption{Performance comparison of PNLMS with Reformulated PLMS}
\label{plms_lms}
\end{figure}

Unlike the NLMS, the normalization in the denominator of PNLMS
update is a weighted normalization of the input vector and the weight
matrix (G) changes in each iteration. So we can't compute the
normalization term recursively and calculating this vector matrix
vector product freshly in each iteration requires 2L
multiplications, L additions and adds significant area and time
complexity (complexity grows with L, L would be generally large for real-time applications) to the already complex proportionate adaptation. To alleviate this problem, we proposed PLMS $\cite{vinay}$ (which is similar to LMS) by removing this weighted normalization and analyzed the convergence performance and it is shown that the PLMS algorithm is stable under $0 < \mu < 2$ for a white input with mean zero and unit variance and is able to perform as the original PNLMS. The  PLMS update equation is given by,
\vspace{0.5em}
\begin{equation}
\textbf{w}(n+1)=\textbf{w}(n)+\mu \hspace{0.5mm} {\textbf{G}}(n)\hspace{0.5mm} \textbf{u}(n) \hspace{0.5mm} e(n),
\end{equation}
where gain matrix calculation is same as that of simplified PNLMS
algorithm. Now, the performance of the re-formulated
algorithm is compared (with simplified gain calculation and without
normalization) with that of the original PNLMS. For this, we considered a
sparse system identification problem and we use the the sparseness measure $S_{m} = \frac{L}{L- \sqrt{L}} \big (1- \frac{{\parallel h \parallel}_{1}}{\sqrt{L} {\parallel h \parallel}_{2}} \big )$ to characterize the sparse system. The unknown system of length $512$ of which only $64$ taps are active ($S_{m} = 0.8637$) is considered.
Learning curves (Normalized MSD vs Iterations) shown in
Fig.~\ref{plms_lms} are obtained by averaging over $500$
experiments. All the other simulation parameters are shown in the text box. As we can see PNLMS and reformulated PLMS outperforms
the conventional NLMS significantly. We can also see that the
performances of reformulated PLMS and PNLMS are same and there is no
penalty for the proposed reformulations.

\subsection{Delayed Proportionate LMS (DPLMS)}
The LMS algorithm in its original form is not suitable for realtime
VLSI implementations because of the coefficient update feedback
loop. To overcome this, LMS is modified to a form known as delayed
LMS (DLMS) $\cite{dlms1}$-$\cite{dlms2}$ with an assumption that the error
gradient $e(n)*{\bf u}(n)$ does not change much with the delay
$M$. The weight update equation of DLMS is shown below:
\begin{equation}\label{dlms_w}
\begin{split}
{\bf w}(n+1)={\bf w}(n) + \mu \hspace{0.5mm}  {\bf u}(n-M) \hspace{0.5mm} e(n-M),
\end{split}
\end{equation}
where $e(n-M)=d(n-M) - y(n-M)$. With the introduction of the $M$
delay registers in the feedback loop, we can apply re-timing
$\cite{retime}$ to the LMS circuit, so that the critical path is reduced to either one multiplier $T_{mult}$ or one
adder $T_{add}$ as desired by the application. The concept of
delayed adaptation can be extended to proportionate LMS also and the
corresponding weight update equation for the delayed PLMS (DPLMS)
algorithm is given by,
\begin{equation}
\textbf{w}(n+1)=\textbf{w}(n)+\mu {\textbf{G}(n-M)\textbf{u}(n-M)e(n-M)},
\end{equation}
where
\begin{equation} \textbf{G}(n-M)=diag(g_{0}(n-M),g_{1}(n-M),...g_{L-1}(n-M)),
\end{equation}
with
\begin{equation}
g_{i}(n-M)=\frac{\gamma_{i}(n-M)}{\sum\limits_{i=0}^{L-1}\gamma_{i}(n-M)}, \mspace{8mu} 0\leq i \leq (L-1),
\end{equation}
and
\begin{equation}
\gamma_{i}(n-M)=
\mathrm{F}[(w_{i}(n-M) +\rho)],
\end{equation}

\begin{equation}
\mathrm{F}[(w_{i}(n-M) = |w_{i}(n-M)|.
\end{equation}

\begin{figure}[h!]
\centering
\includegraphics [height=70mm,width=75mm]{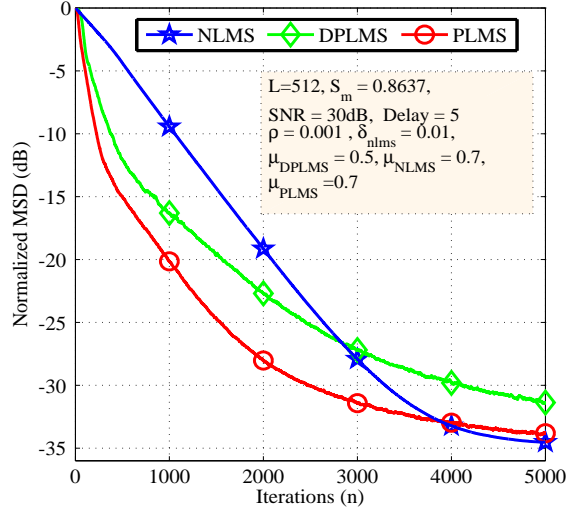}
\caption{Performance comparison of PLMS with Delayed PLMS}
\label{dlms_lms}
\end{figure}

Now we compare the performance of the DPLMS with that of PLMS. We
consider the same sparse system identification problem as in last subsection and
simulation parameters are shown in the text box. Learning curves
(Normalized MSD vs Iterations) shown in Fig.~\ref{dlms_lms} are
obtained by averaging over $500$ experiments. We can notice that
after the initial phase, the convergence rate of the PLMS algorithm
slows down significantly, even becoming slower than NLMS. The large
coefficients converge very fast at the cost of slowing down
dramatically convergence of the small coefficients. DPLMS suffers
even more because of the direct proportional gain and delayed
adaptation. To address this issue, these re-formulations are extended to the $\mu$-Law PLMS (MPLMS) which has more balanced gain distribution.

\subsection{Delayed $\mu$-Law PLMS (DMPLMS)}

\begin{figure}[h!]
\centering
\includegraphics [height=70mm,width=75mm]{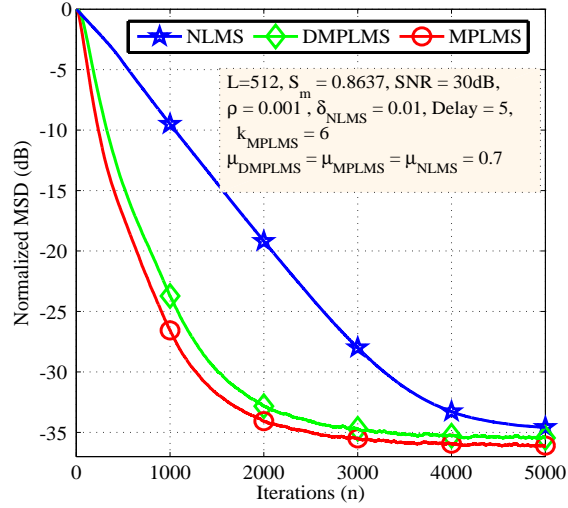}
\caption{Performance comparison of MPLMS with Delayed MPLMS}
\label{dmplms_mplms}
\end{figure}

$\mu$-Law PNLMS (MPNLMS) $\cite{mplms}$ offers more balanced distribution
of the adaptation energy among all the coefficients as it is based on
the optimization criteria that all the coefficients converge
simultaneously to the $\epsilon$-vicinity of their true value so that the overall convergence is the fastest. Here
gain is proportional to the natural logarithm of the absolute value of the current weight instead of the absolute of current weight itself. This logarithm in the $\mathrm{F}[(.)]$ equation (which is shown below) is the only difference between MPNLMS and PNLMS. The parameter $\xi$ is a very small positive number and its value should be chosen based on the measurement noise level. For applications such as network echo cancelation, $\xi = 0.001$ is a good choice because the echo below $- 60 dBm$ is negligible.
\begin{equation}
\mathrm{F}[(w_{0}(n))] = \ln(1+ \frac{|w_{0}(n)|} {\xi}).
\end{equation}
We extended all the aforementioned re-formulations to MPNLMS and also we simplified the $\mathrm{F}[(.)]$ equation to the following equation for our VLSI implementations and verified its effectiveness through simulations. Since we changed the natural logarithm to base-2 logarithm, $k$ is chosen to be $6$.
\begin{equation}
\mathrm{F}[(w_{0}(n))] = \log_{2}(1+ \frac{|w_{0}(n)|} {2^{-k}}).
\end{equation}
Because of this balanced gain distribution delayed adaptation of
MPLMS doesn't suffer much and performances of MPLMS and DMPLMS (for
same experimental conditions) are shown in Fig.~\ref{dmplms_mplms}.
We can see that delayed MPLMS with logarithmic simplifications is a
robust algorithm for white input case which is suitable for VLSI implementation. The formal algorithm is
summarized in Algorithm 1.

\begin{figure}[h!]
\centering
\includegraphics [height= 100mm,width=80mm]{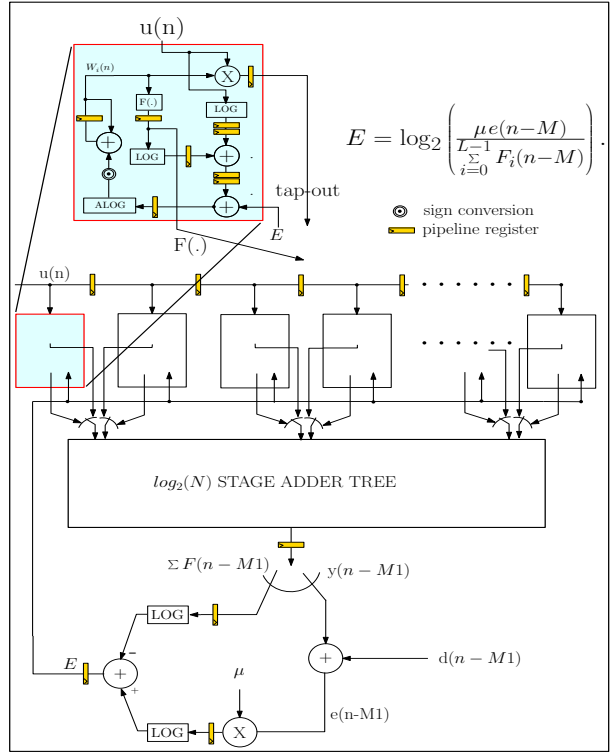}
\caption{MPLMS Architecture}
\label{mplms_arch}
\end{figure}

\begin{algorithm}
\DontPrintSemicolon 
Initialization : $ w_{i}(1)= 0, \mspace{8mu} 0\leq i \leq (L-1)$ \;
Parameters : $\mu$, $\rho$, $k$ \;

Updation :\;
 \vspace{0.5em}
 $e(n) = d(n)-\textbf{w}^{T}(n)
\textbf{u}(n)$\; $\mathrm{F}[(w_{i}(n))] = \log_{2}(1+
\frac{|w_{i}(n)|} {2^{-k}})$\;
 $\gamma_{i}(n)=
\mathrm{F}[(w_{i}(n) +\rho)]$\;
$g_{i}(n)=\frac{\gamma_{i}(n)}{\sum\limits_{i=0}^{L-1}\gamma_{i}(n)},\mspace{8mu}
0\leq i \leq (L-1)$\;
 $\textbf{G}(n)$ =
$diag(g_{0}(n),g_{1}(n),...g_{L-1}(n))$ \;
$\Delta \textbf{W} = \mu
{\textbf{G}(n-M)\textbf{u}(n-M)e(n-M)}$\;
$\textbf{w}(n+1)=\textbf{w}(n)+\Delta \textbf{W}$\;
\caption{Delayed MPLMS algorithm} 
\label{llad_e_comp}
\end{algorithm}
\subsection{Logarithmic Number System to calculate the gradient}
Even the simplified DPLMS gradient calculation involves division and
fixed point division is quite a bit more complicated $\cite{div_cmp}$ than fixed
point multiplication, and usually takes a lot more cycles than
performing a multiplication. Thus we use logarithmic number system
(LNS) which simplifies the division. The logarithmic conversion of a real number $Q$ can be obtained by  $\log_{2}(Q) = k+ \log_{2}(1+x)$  when $Q =  2^{k}(1+x)$, $k$ is the leading one bit position and $x$ is a fraction. Depending on how we approximate this $\log_{2}(1+x)$ there have been multiple schemes in the literature. This can be achieved by using a LUT with $x$ as index and $\log_{2}(1+x)$ as the value or performing simple conversion on $x$. Mitchell $\cite{mitchell}$ used the value of $x$ to directly approximate $\log_{2}(1+x)$. In this scheme the log-converter contains a simple Leading One Detector (LOD) circuit followed by a barrel shifter and antilog converter is a concatenation of $1$ and $x$ followed by a barrel shifter. DMPLMS architecture employing these log/antilog converters will be explained in the next section.

\subsection{Proposed Architecture}
The architecture for the proposed DMPLMS and DPLMS is shown in Fig.~\ref{mplms_arch}. Please note that the only difference between DPLMS and DMPLMS architectures is the $F(.)$ module.
One of the PLMS/MPLMS tap is zoomed and showed separately. Each tap gets two inputs one is the regressor input from the tapped delay line,
the other is the quantity E, which is defined as,
\begin{equation}
E=  \log_{2} \left( \frac{\mu \hspace{0.5mm} e(n-M)}{\sum\limits_{i=0}^{L-1}F_{i}(n-M)} \right).
\label{e_eq}
\end{equation}
and it is used in the gradient calculation. As mentioned in previous subsection, to reduce the complexity
we use the LNS for gradient calculation. Since logarithm of a negative number is undefined,
we take the logarithm of absolute value of $e(n-M)$ and propagate the sign of $e(n-M)$ to the sign conversion unit in each tap.
Please note that sign of this E is same as sign of $e(n-M)$ as the denominator of Equation \eqref{e_eq} is always positive.
Gradient in LNS format can be written as
\begin{equation}
\Delta \textbf{W} = 2^{\log_{2}(\mu \hspace{0.5mm}
{\textbf{G}}(n-M)\hspace{0.5mm} \textbf{u}(n-M)\hspace{0.5mm} e(n-M))} .
\end{equation}
By expanding the gain matrix and by separating the terms which are independent of tap and  which are
dependent on tap, we get
\begin{equation}
\Delta \textbf{W} = 2^{\log_{2} \left(\frac{\mu \hspace{0.5mm}
 |e(n-M)|}{\sum\limits_{i=0}^{L-1}\mathrm{F}[w_{i}(n-M)]}\right)+ \log_{2}(
{\mathrm{F}[|w_{i}(n-M)|] |\textbf{u}(n-M)})|}.
\end{equation}
By replacing the first term with E which was defined above, we get the final
equation for the gradient
\begin{equation}
\Delta \textbf{W} = 2^{E + \log_{2}(\mathrm{F}[w_{i}(n-M)]) +
\log_{2} (|\textbf{u}(n-M)|)} .
\end{equation}
As shown in each tap, logarithms of $\mathrm{F}[|w_{i}(n)|])$ and
$|u_{i}(n)|$  are added and delayed by the required amount to match
the delay in the filter output path and the sum is then added to $E$, finally antilogarithm is applied to the resultant to get the magnitude of the gradient. Sign of the gradient is decided by XOR operation between sign$(e(n-M))$ and sign$(u(n-M))$.
After the required sign conversion, the result is added to the old weight to get the new weight for each tap.
Each tap generates two outputs namely $Tap\_out$ and $\mathrm{F}[w_{i}(n)]$.
These two quantities of all the taps need to be added using an adder
tree structure to calculate the final filter output and the
denominator of the gain matrix respectively. So we fold the adder
tree by a folding factor of $2$ i.e., use the same adder tree in a
time shared fashion to calculate both these quantities. If we use a carry save adder tree, the propagation delay would be less and thus we can run the adder tree at twice the clock rate compared to the FIR filter.

 \subsection{Fixed point implementation}
\begin{figure}[h!]
\centering
\includegraphics [height=70mm,width= 75mm]{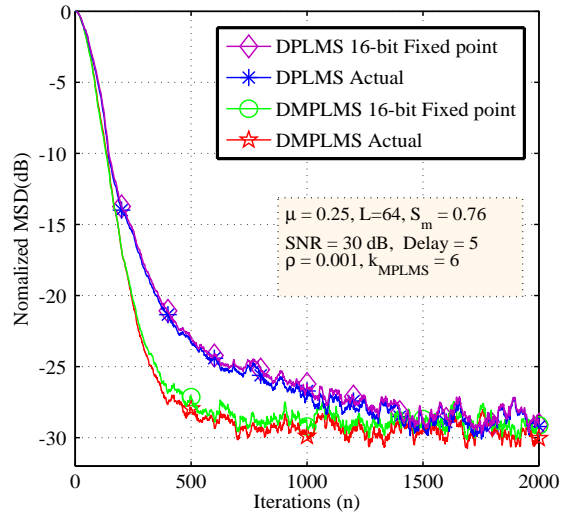}
\caption{Performance of fixed point implementation}
\label{plms_fix}
\end{figure}

We have done simulations using MATLAB with $16$-bit fixed point representations and with all the above suggested re-formulations and compared the results with those of floating point simulations to see if the algorithms are tolerant to the re-formulations and LNS approximations. Fig.~\ref{plms_fix} shows the comparison of both DPLMS and DMPLMS algorithms learning curves of MSD  by averaging over $10$ runs. Because of the iterative and stochastic nature of the algorithms, we can notice that they are tolerant to the logarithmic approximations and hence $16$-bit fixed point and floating point curves coincide.
\begin{figure*}[t!]
\centering \subfigure[Sparse Impulse Response. \label{S_imp}]{
\includegraphics[height=27mm,width=58mm]{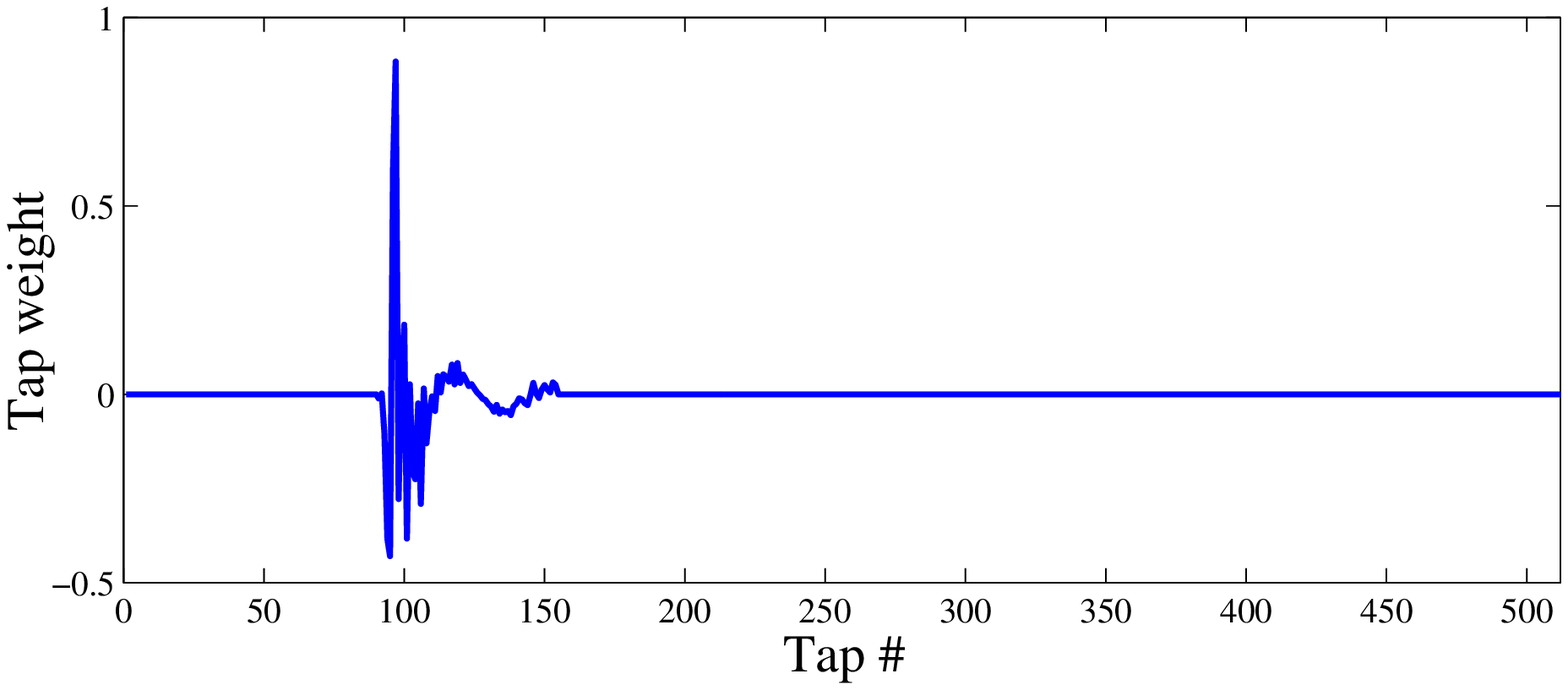} }
\hspace{-5mm} \subfigure[DCT Weights.
\label{dct_imp}]{
\includegraphics[height=27mm,width=58mm]{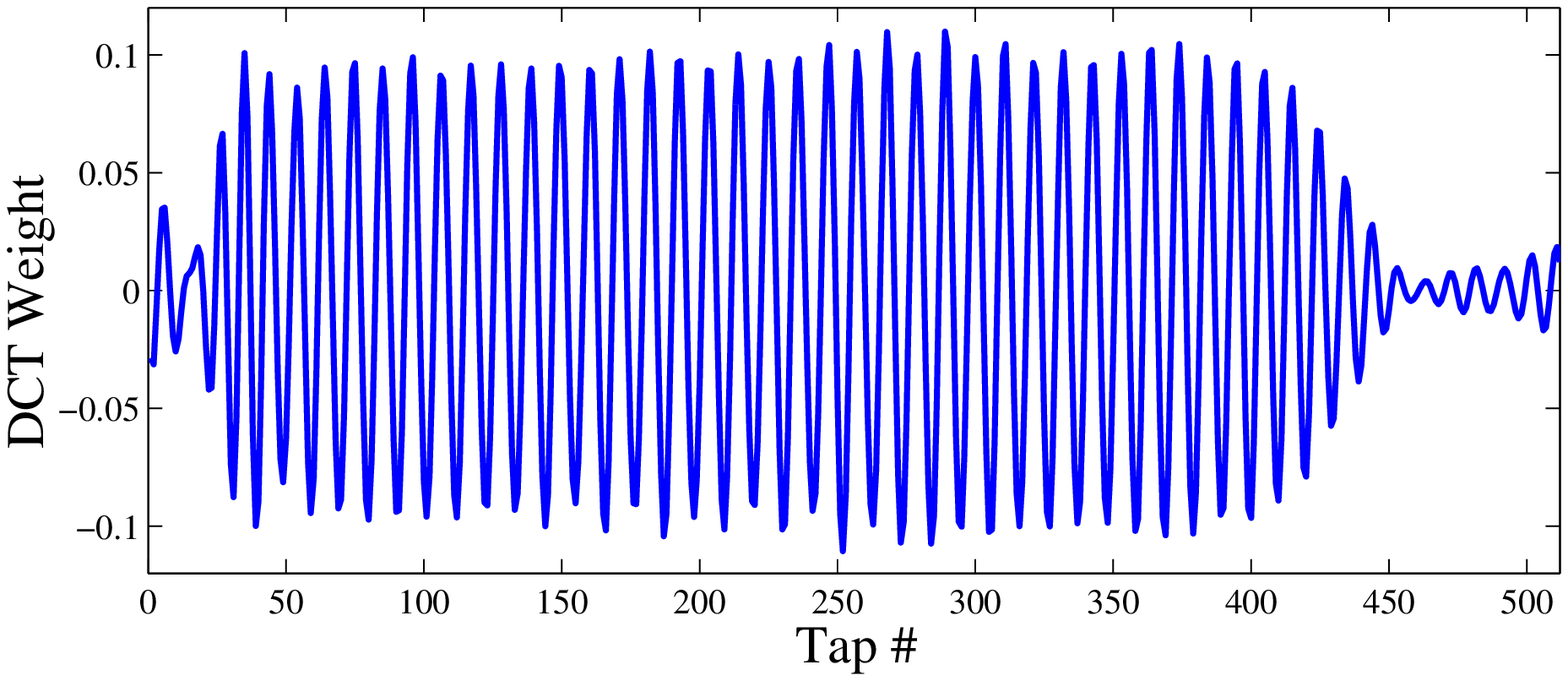} }
\hspace{-5mm} \subfigure[DWT Weights.
\label{dwt_imp}]{
\includegraphics[height=27mm,width=58mm]{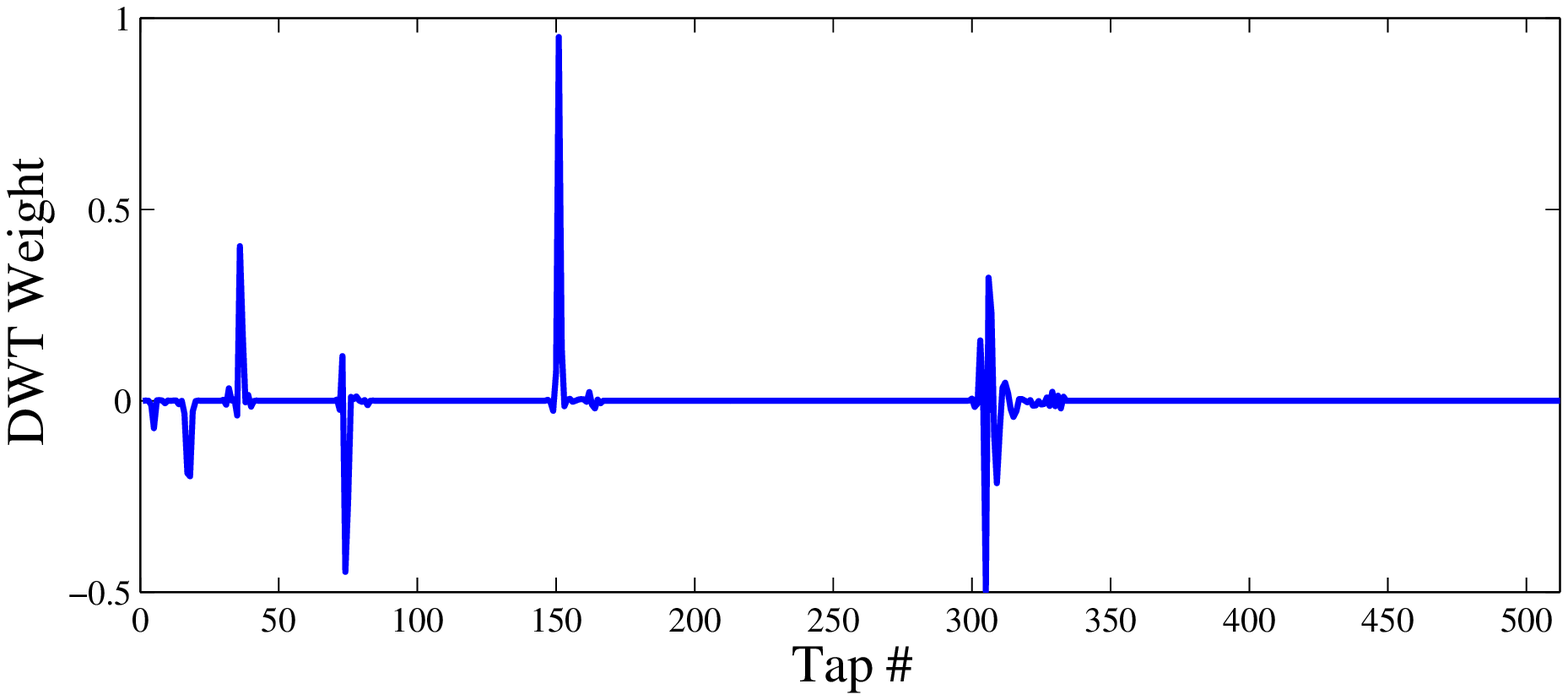} }
\vspace{-2mm} \caption{Comparison of DCT/DWT coefficients.
} \label{dct_wave}
\end{figure*}

\section{Transform Domain Delayed Wavelet MPLMS Algorithm}
\subsection{Motivation}
\begin{figure}[h!]
\centering
\includegraphics [height=70mm,width= 80mm]{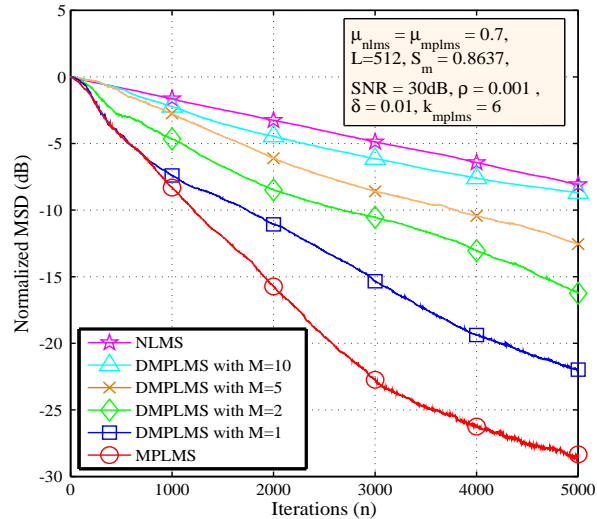}
\caption{Performance degradation of DMPLMS for color input}
\label{dmpl_diverge}
\end{figure}

The convergence performance of LMS-type filters is highly dependent
on the correlation of the input data and, in particular, on the
eigenvalue spread of the input correlation matrix $\bf{R}$.
Because of this, the convergence of LMS-type filters becomes
prohibitively slow for correlated input such as speech. We also found that the effect
of delayed adaptation on the convergence is much more pronounced in
the case of colored input compared to the white input, which can be observed from Fig.~\ref{dmpl_diverge},

It can also be noted that the performance degrades with the adaptation delay $M$ for the DMPLMS algorithm and becomes worse than that of NLMS when the adaptation delay is increased beyond $10$ clocks. This problem can be addressed by transform domain adaptive filters $\cite{td_adf}$. The main idea behind transform domain adaptive filters is that a
de-correlating unitary transform is applied to the input and then by
power normalizing each input component in the transform domain, we
can make the autocorrelation matrix approximately an identity
matrix. This approximate white input is then fed to the adaptive
filter for adaptation. However, please note that now the estimated filter weights are also in transform domain.
Hence we need to select the transform which de-correlates the input while preserving the sparsity of the transformed filter weights.
For example Discrete Cosine Transform (DCT) $\cite{td_adf}$ which has excellent de-correlating properties
but doesn't preserve the sparsity which can be observed from Fig.~\ref{S_imp} and Fig.~\ref{dct_imp},
and thus we will loose the advantage of proportional adaptation. On the other hand, Discrete Wavelet Transform (DWT) with its time frequency localization property,
preserves the sparseness of impulse response $\cite{k_c_ho}$ as shown in Fig.\ref{dwt_imp} and also exhibits decent de-correlation property. Thus Wavelet MPNLMS (WMPNLMS) $\cite{deng_wave}$ is more suitable for sparse adaptive filters under correlated input. Hence, before feeding the input to the proposed DMPLMS adaptive filter, it will be de-correlated using DWT. The resultant DWMPLMS algorithm with all the proposed re-formulations is summarized in Algorithm 2.
\begin{algorithm}
\DontPrintSemicolon 
Initialization : $ w_{i}(1)= 0, \mspace{8mu} 0\leq i \leq (L-1)$ \;
Parameters : $\mu$, $\rho$, $\beta$, $k$ \;

Updation :\;

 $\textbf{u}_T(n) = \textbf{T} \textbf{u}(n)$ \;
 $\Psi_{i}(n) = \beta \Psi_{i}(n-1) + (1-\beta)
 |(u_{T,i}(n))| , \mspace{8mu} 0\leq i \leq (L-1)$\;
 $\textbf{D}(n)$ =
  diag($\Psi_{i}(n)$) \;
 $e(n) = d(n)-\textbf{w}_{T}^{T}(n)
\textbf{u}_{T}(n)$\; $\mathrm{F}[(w_{T,i}(n))] = \log_{2}(1+
\frac{|w_{T,i}(n)|} {2^{-k}})$\;
 $\gamma_{i}(n)=
\mathrm{F}[(w_{T,i}(n))+\rho]$\;
$g_{i}(n)=\frac{\gamma_{i}(n)}{\sum\limits_{i=0}^{L-1}\gamma_{i}(n)},\mspace{8mu}
0\leq i \leq (L-1)$\;
 $\textbf{G}(n)$ =
$diag(g_{0}(n),g_{1}(n),...g_{L-1}(n))$ \;
$\Delta \textbf{w} = \mu {\textbf{G}(n-M)
\textbf{D}^{-1}(n-M) \textbf{u}(n-M)e(n-M)}$\;
$\textbf{w}_{T}(n+1)=\textbf{w}_{T}(n)+ \Delta \textbf{w}$\;
\caption{Delayed WMPLMS algorithm} 
\label{dwmpls_algo}
\end{algorithm}

In the next section we study different wavelet transforms and their convergence performances in the framework of delayed proportionate adaptation.

\subsection{Study of various wavelet Transforms and their de-correlating properties}

\begin{figure}[h]
\centering
\includegraphics [height=70mm,width= 80mm]{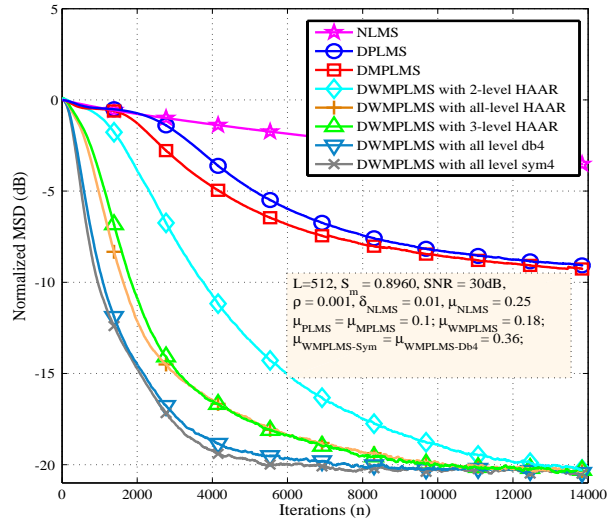}
\caption{Performance of 3-level HAAR DWMPLMS}
\label{3_haar_MSD}
\end{figure}

We considered three families of wavelets  $\cite{strang}$ namely Haar, Symlets denoted as Sym4 and Daubechies denoted as db4
for the comparative study of their convergence performance when used
in DWMPLMS algorithm and the results are shown in
Fig.~\ref{3_haar_MSD}. We see that DWMPLMS algorithms outperforms
DMPLMS and DPLMS algorithms significantly. Among the wavelet families
Sym4 perform better than the other two. Even though Haar
wavelet's performance is inferior to the other two, when we consider
the performance complexity tradeoff as a metric, we show that Haar
is the better than the rest. Next we analyzed the performance with various levels of Haar
decomposition and we noticed that there is a significant performance
improvement from two to three levels and beyond three levels of decomposition the
performance improvement is negligible. Thus we considered $3$-level
Haar wavelet for implementing DWMPLMS algorithm. In the next section
we show that how a fast sliding wavelet transform can be computed by
taking the streaming nature of the input data into account.

\subsection{Sliding Wavelet transform}
\begin{figure*}[t!]
\centering \subfigure[DWMPLMS Architecture.
\label{dwmpls_arch}]{
\includegraphics[height=110mm,width=84mm]{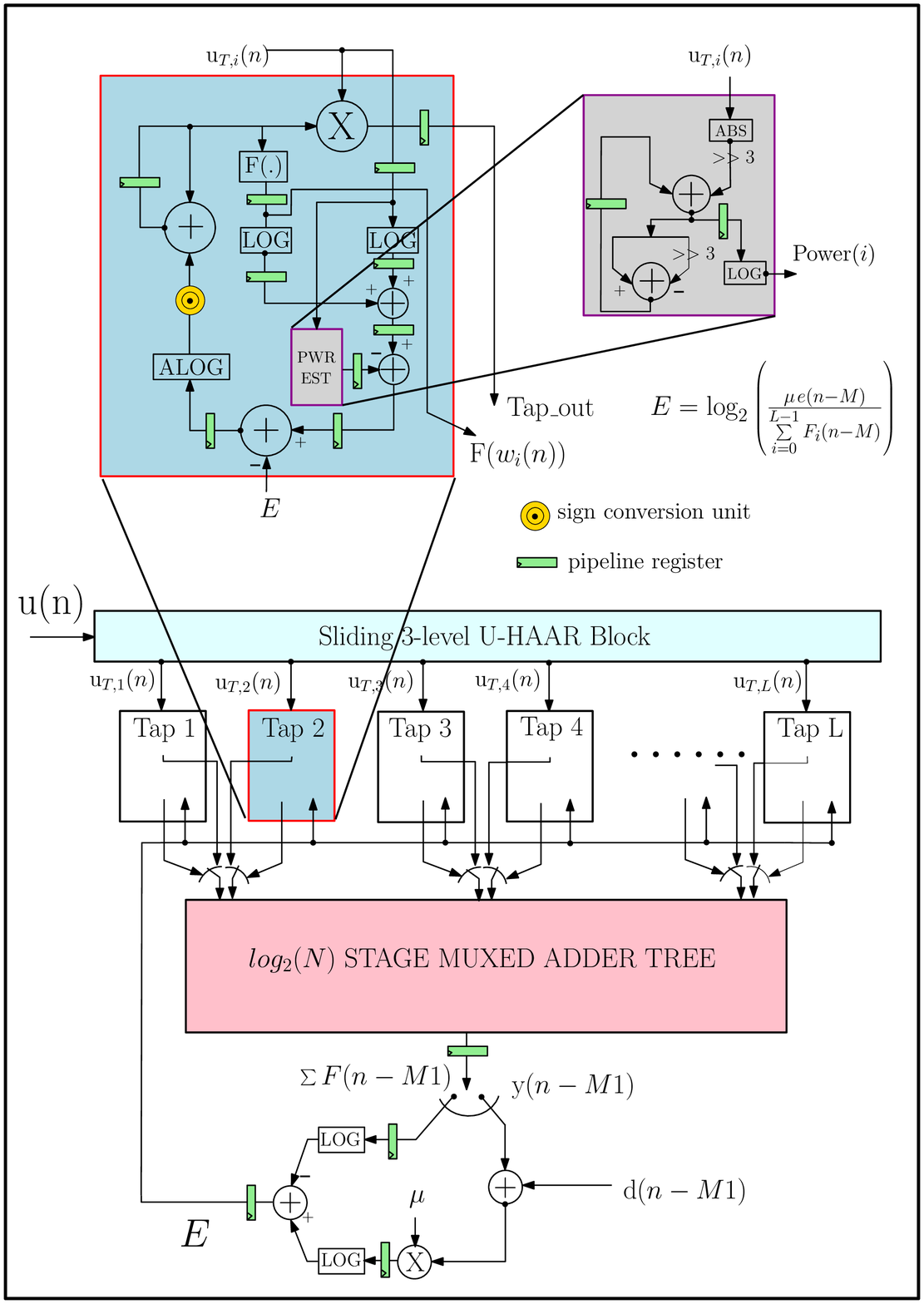} }
\subfigure[Sliding HAAR Architecture.
\label{3_haar}]{
\includegraphics[height=110mm,width=84mm]{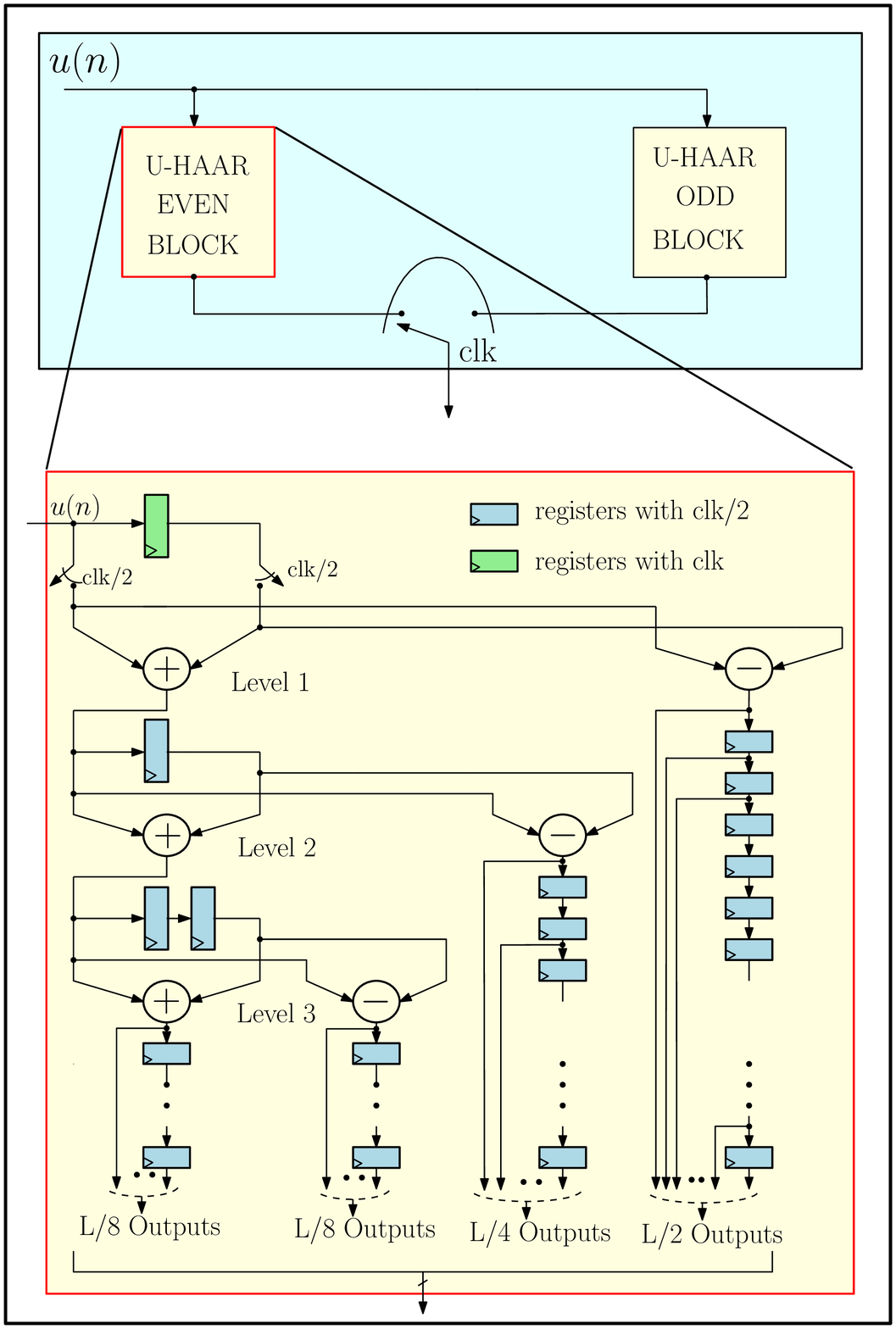} }
 \caption{Proposed Architectures.
} \label{System_Response}
\end{figure*}

The data vector $\textbf{u}(n)$ (which is defined as $\textbf{u}(n)
= [u(n) u(n-1) ... u(n-N+1)]^{T}$) is updated at each new iteration by letting one data sample to enter
in and one to leave. The transformed input vector is denoted by $\textbf{u}_T(n)$ i.e.
$\textbf{u}_T(n) = \textbf{T} \textbf{u}(n)$. The streaming nature of the input can be used to exploit the
redundancies that exist between calculation of running wavelet
transforms of $\textbf{u}(n)$ and $\textbf{u}(n+2)$ (where $\textbf{u}(n+2)
= [u(n+2) u(n+1) ... u(n-N+3)]^{T}$). Let $\textbf{T}_{8}$ be 8-point $Sym2$ DWT which has four
low frequency coefficients $h0, h1, h2$ and $h3$ and four high frequency coefficients $g0, g1,
g2$ and $g3$. Now consider the matrix vector products $\textbf{T}_{8} \textbf{u}(n)$ which is shown below and $\textbf{T}_{8}
\textbf{u}(n+2)$ for an $8$-tap filter
\[
\hspace{-8mm}
    \begin{pmatrix}  h0 & h1 & h2 & h3 & 0 & 0 & 0 & 0 \\
    0 & 0 & h0 & h1 & h2 & h3 & 0 & 0 \\
    0 & 0 & 0 & 0 & h0 & h1 & h2 & h3 \\
    h2 & h3 & 0 & 0 & 0 & 0 & h0 & h1 \\
    g0 & g1 & g2 & g3 & 0 & 0 & 0 & 0 \\
    0 & 0 & g0 & g1 & g2 & g3 & 0 & 0 \\
    0 & 0 & 0 & 0 & g01 & g1 & g2 & g3 \\
    g2 & g3 & 0 & 0 & 0 & 0 & g0 & g1  \end{pmatrix}
  \begin{pmatrix} u(n) \\
    u(n-1) \\
    u(n-2) \\
    u(n-3) \\
    u(n-4) \\
    u(n-5) \\
    u(n-6) \\
    u(n-7)  \end{pmatrix}.
\]
 In $\cite{dwt_slide}$, it has been shown that by defining basic adder cells and reusing the partial sums, computational complexity  can be significantly reduced and the number of multiplications can even become independent of the filter order under certain conditions. However, this reduction is true only for one level of decomposition. Once we go for second level of decomposition,
  i.e., $\left(
  \begin{matrix}  \textbf{T}_{4} & \textbf{0}_{4} \\
        \textbf{0}_{4} &  \textbf{I}_{4} \end{matrix}\right)
  \textbf{T}_{8}\textbf{u}(n)
$, the redundancies cease to exist since the wavelet coefficients obtained in the first level, i.e., $\textbf{T}_{8}\textbf{u}(n)$ no longer have the streaming nature.

 This complicates the wavelet transform computation and thus not
suitable for DWMPLMS algorithm where we need at least 3-levels of
decomposition to get good de-correlating properties. On the other hand if we
consider the un-normalized Haar wavelet whose coefficients are simple $\pm
1$s, the redundancies continue to exist through multiple levels of
decomposition and we can exploit them using a regular structure. For example consider 8-point un-normalized
HAAR wavelet matrix with two levels of decomposition (i.e., the product matrix $\begin{pmatrix}  \textbf{T}_{4} & \textbf{0}_{4} \\
        \textbf{0}_{4} &  \textbf{I}_{4} \end{pmatrix}
  \textbf{T}_{8} $ )  which is shown below:
\[
  \begin{pmatrix}  1 & 1 & 1 & 1 & 0 & 0 & 0 & 0 \\
    0 & 0 & 0 & 0 & 1 & 1 & 1 & 1 \\
    1 & 1 & -1 & -1 & 0 & 0 & 0 & 0 \\
    0 & 0 & 0 & 0 & 1 & 1 & -1 & -1 \\
    1 & -1 & 0 & 0 & 0 & 0 & 0 & 0 \\
    0 & 0 & 1 & -1 & 0 & 0 & 0 & 0 \\
    0 & 0 & 0 & 0 & 1 & -1 & 0 & 0 \\
     0 & 0 & 0 & 0 & 0 & 0 & 1 & -1  \end{pmatrix}
  \begin{pmatrix} u(n) \\
    u(n-1) \\
    u(n-2) \\
    u(n-3) \\
    u(n-4) \\
    u(n-5) \\
    u(n-6) \\
    u(n-7)  \end{pmatrix}.
\]
Now, consider the low frequency wavelet components at $n^{th}$ time index \;
  \begin{equation}
  \begin{split}
u_{T,0}(n) &= u(n) + u(n-1) + u(n-2) + u(n-3), \\
u_{T,1}(n) &= u(n-4) + u(n-5) + u(n-6) + u(n-7), \\
u_{T,2}(n) &= u(n) + u(n-1) - [u(n-2) + u(n-3)], \\
u_{T,3}(n) &= u(n-4) + u(n-5) -[u(n-6) + u(n-7)], \\
\end{split}
\end{equation}

if we define $a(n) = u(n) + u(n-1)$, then $u_{T,0}(n) = a(n) + a(n-2)$ and  $u_{T,2}(n) = a(n) - a(n-2)$. Similarly at $(n+2)^{th}$ time index \;
  \begin{equation}
  \begin{split}
u_{T,0}(n+2)    &= u(n+2) + u(n+1) + u(n) + u(n-1),\\
u_{T,1}(n+2)   &= u(n-2) + u(n-3) + u(n-4) + u(n-5),\\
u_{T,2}(n+2)   &= u(n+2) + u(n+1) -[u(n) + u(n-1)],\\
u_{T,3}(n+2)   &= u(n-2) + u(n-3) -[u(n-4) + u(n-5)].\\
  \end{split}
  \end{equation}

Now, $u_{T,0}(n+2) = a(n+2) + a(n)$ and  $u_{T,2}(n+2) = a(n+2) - a(n)$. Note that only $a(n+2)$ need to be evaluated at $(n+2)^{th}$ time index and all the other partial sums can be reused. Similarly high frequency components at $n^{th}$ time index \;
  \begin{equation}
  \begin{split}
 u_{T,4}(n)  &= u(n) - u(n-1), \\
 u_{T,5}(n) &= u(n-2) - u(n-3), \\
 u_{T,6}(n)  &= u(n-4) - u(n-5), \\
 u_{T,7}(n)  &= u(n-6) - u(n-7), \\
   \end{split}
 \end{equation}
  and at $(n+2)^{th}$ time index \;
 \begin{equation}
  \begin{split}
 u_{T,0}(n+2)  &= u(n+2) - u(n+1), \\
 u_{T,0}(n+2)  &= u(n) - u(n-1), \\
 u_{T,0}(n+2)  &= u(n-2) - u(n-3, \\
 u_{T,0}(n+2)  &= u(n-4) - u(n-5). \\
   \end{split}
  \end{equation}
Once two input samples are subtracted, the resultant is passed through a tapped delay line and $\frac{L}{2}$ outputs are taken from $\frac{L}{2}$ registers. Similar redundancies exist between $(n-1)^{th}$ and $(n+1)^{th}$ time indices. This scheme can be extended to multiple levels of decomposition using only adders/subtractors and registers. However, please note that un-normalized Haar wavelet is non-orthogonal and this creates issue while computing the filter output, which need to be addressed and it is explained later. The architecture of DWMPLMS with sliding wavelet implementation is explained in the next section.

\subsection{Proposed Architecture}

DWMPLMS architecture is shown in Fig.~\ref{dwmpls_arch} and is similar to DMPLMS Architecture except for the sliding wavelet transform block,
muxed adder tree and power normalization block in each tap. These changes will be explained in detail. In this subsection we explain the
sliding Haar wavelet block and in subsequent sections we explain the others.
Sliding un-normalized Haar wavelet implementation is shown in the Fig.~\ref{3_haar}.
It contains two identical blocks U-HAAR even and U-HAAR odd blocks. U-HAAR even block is zoomed and shown separately.
The switches shown in this U-HAAR even block closes at the rate of $\frac{clk}{2}$ so as to sample only
non-overlapped even pairs like $u(n) , u(n-1)$ and $u(n+2) , u(n+1)$. Similarly the odd pairs $u(n+1) , u(n)$ and $u(n+3) , u(n+2)$  are sampled in U-HAAR odd block.
Once the even pair is received, it is added and subtracted to get the low and high frequency components and high frequency components are passed through a series of registers which are clocked on $\frac{clk}{2}$ which forms the tapped delay line and the $\frac{L}{2}$ high frequency outputs come from this delay line.
The low frequency components are passed through the second level of decomposition. For second level difference components, valid outputs are generated once in four clocks.
Hence we require two registers which run on $\frac{clk}{2}$. If we use one register which run on $\frac{clk}{4}$,
we loose the intermediate results and hence output would be erroneous. Except the top latch shown in green color, all other latches run on $\frac{clk}{2}$. Similarly in the third level,
we need four registers running on $\frac{clk}{2}$ between valid outputs to capture all the intermediate results. In the first level, we need $\frac{(L-2)}{2}$ registers for the difference. Similarly at the second stage we need $\frac{(L-4)}{2}$ registers and at the final stage we need $\frac{(L-8)}{2}$ registers for detail coefficients and another $\frac{(L-8)}{2}$ registers for average coefficients. We see that it is a register hungry design but the critical path is just three adders/subtractors since there are no multipliers. Similarly the redundancies between $\bf{u}(n+1)$ and $\bf{u}(n-1)$ are exploited in the U-HAAR ODD block. U-HAAR ODD block also has a same structure and also it runs on $\frac{clk}{2}$ but in the complementary phase of U-HAAR EVEN block clock. At the output there is a switch which touches one of the sides depending the even/odd number of clock and the wavelet components are fed to the adaptive filter.
We can observe that the critical path for both DMPLMS and DWMPLS architectures is $T_{mult}$ i.e. delay of one multiplier.
\subsection{Un-normalized Haar Transform}
The filter output of transform domain adaptive filters is given below. Since both filter weights and input are in transform domain and if the transform is orthogonal,
then the filter output remains unchanged as shown.
 \begin{equation}
  \begin{split}
y(n) &= \textbf{w}_{T}^{T}(n) \textbf{u}_{T}(n) \\
     &= (\textbf{T}\textbf{w}(n))^{T} \textbf{T}\textbf{u}(n)  \\
     &= \textbf{w}^{T}(n) (\textbf{T}^{T} \textbf{T}) \textbf{u}(n) \\
     &= \textbf{w}^{T}(n) (\textbf{I}) \textbf{u}(n) \\
     &= \textbf{w}^{T}(n) \textbf{u}(n).
  \end{split}
 \end{equation}
However, since the Haar wavelet used above is not orthogonal, with
$3$-levels of decomposition $(\textbf{T}^{T} \textbf{T})$ becomes
  \begin{equation}
(\textbf{T}^{T}
\textbf{T}) = \left(
  \begin{matrix}  8 & 0 & 0 & 0 & 0 & 0 & 0 & 0 \\
    0 & 8 & 0 & 0 & 0 & 0 & 0 & 0 \\
    0 & 0 & 4 & 0 & 0 & 0 & 0 & 0 \\
    0 & 0 & 0 & 4 & 0 & 0 & 0 & 0 \\
    0 & 0 & 0 & 0 & 2 & 0 & 0 & 0 \\
    0 & 0 & 0 & 0 & 0 & 2 & 0 & 0 \\
    0 & 0 & 0 & 0 & 0 & 0 & 2 & 0 \\
     0 & 0 & 0 & 0 & 0 & 0 & 0 & 2  \end{matrix},
   \right )
    \end{equation}
   and hence y(n) is given by
\begin{equation}
\begin{split}
\hspace{-4mm} y(n) = 8[(w_{0}u(n)+w_{1}u(n-1)] + 4[w_{2}u(n-2)+w_{3}u(n-3)]  \\ +
          2[w_{4}u(n-4)+ w_{5}u(n-5) + w_{6}u(n-6)+w_{7}u(n-7)].
\end{split}
\end{equation}

\begin{figure}[h!]
\centering
\includegraphics [height= 50mm,width=85mm]{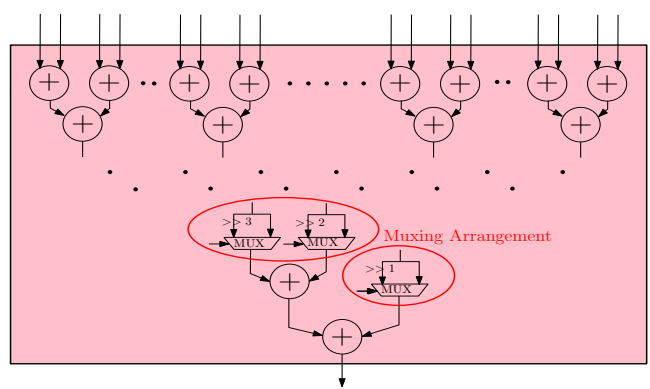}
\caption{Mux Add Tree Architecture}
\label{mux_add_tree}
\end{figure}

In this result, half of the partial MAC (multiply-accumulate)
 terms are scaled up by a factor of $2$, similarly half of the remaining terms are scaled up by a factor of $4$ and remaining terms by $8$. These terms need to be scaled down by the same factors to get the correct output $y(n)$.
 we can achieve this by right shifting the result of the particular adder tree branch as shown in Fig.~\ref{mux_add_tree}.

 Note that this arrangement is independent of the filter length. In the adder tree architecture, the MAC terms which are scaled up by a factor of $2$ always come from second branch of the adder tree
and hence this result is right shifted by one bit position to scale it down by $2$ and this doesn't require any extra hardware..
Similarly the results of other two branches are right shifted by $2$ and $3$ bit positions.
The choice of un-normalized HAAR wavelet has three advantages. We could exploit the redundancies in sliding wavelet
transform through multiple levels of decomposition, we can get away with multipliers in calculating DWT and finally no special hardware required for scaling down the output. Multiplexing arrangement is necessary in the adder tree since we are folding the adder tree by a factor of $2$ (just like the DMPLMS case) and scaling is not needed during calculation of the denominator of proportional gain.

\begin{figure*}[t!]
\centering \subfigure[Highly sparse system (sparse network echo
path in ITU-T G.168), $S_m = 0.8960$. \label{Sparse}]{
\includegraphics[height=27mm,width=58mm]{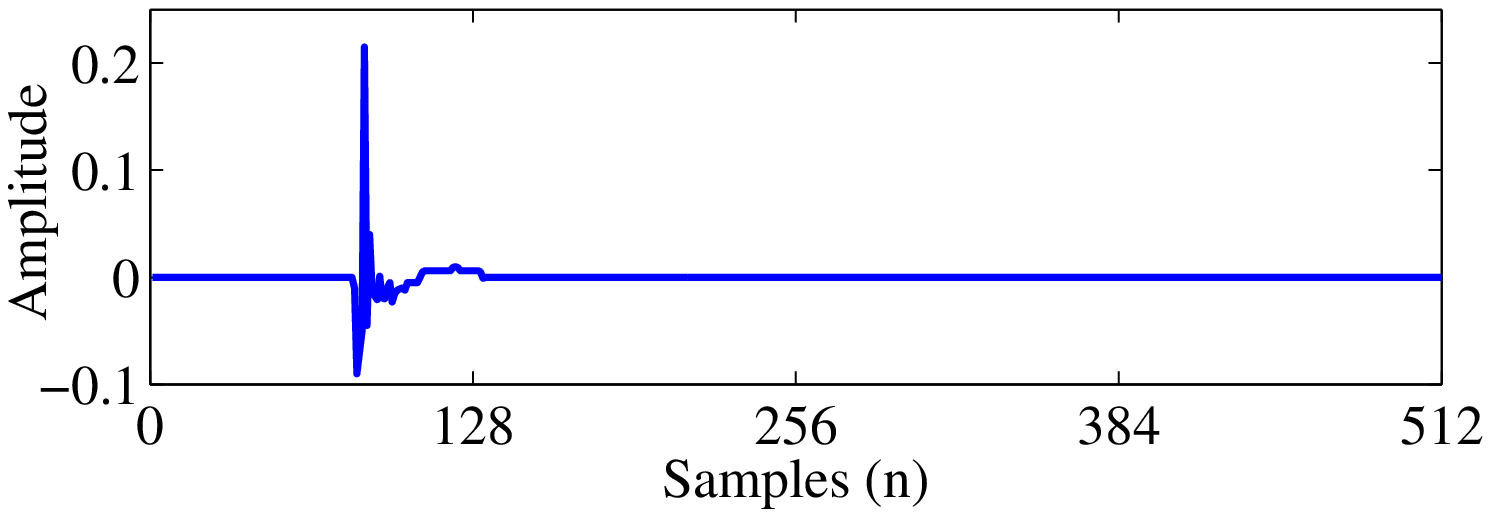} }
\hspace{-5mm} \subfigure[Semi sparse system, $S_m = 0.5560$.
\label{Semi_Sparse}]{
\includegraphics[height=27mm,width=58mm]{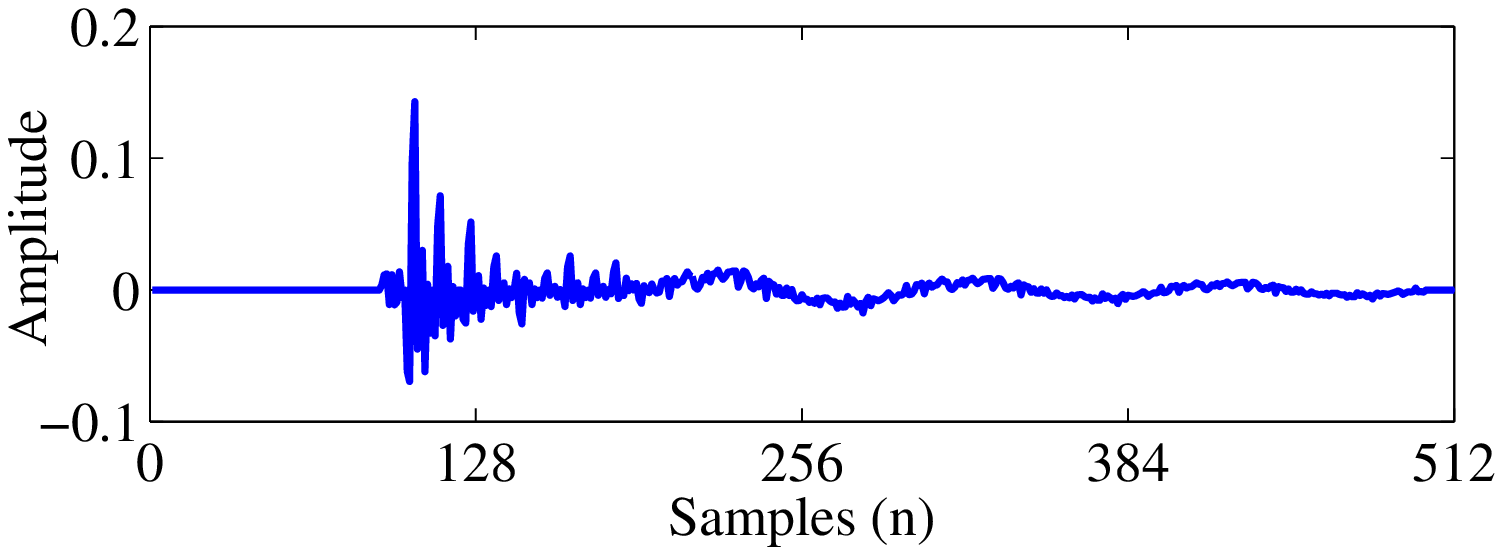} }
\hspace{-5mm} \subfigure[Non-sparse system, $S_m = 0.3486$.
\label{non_Sparse}]{
\includegraphics[height=27mm,width=58mm]{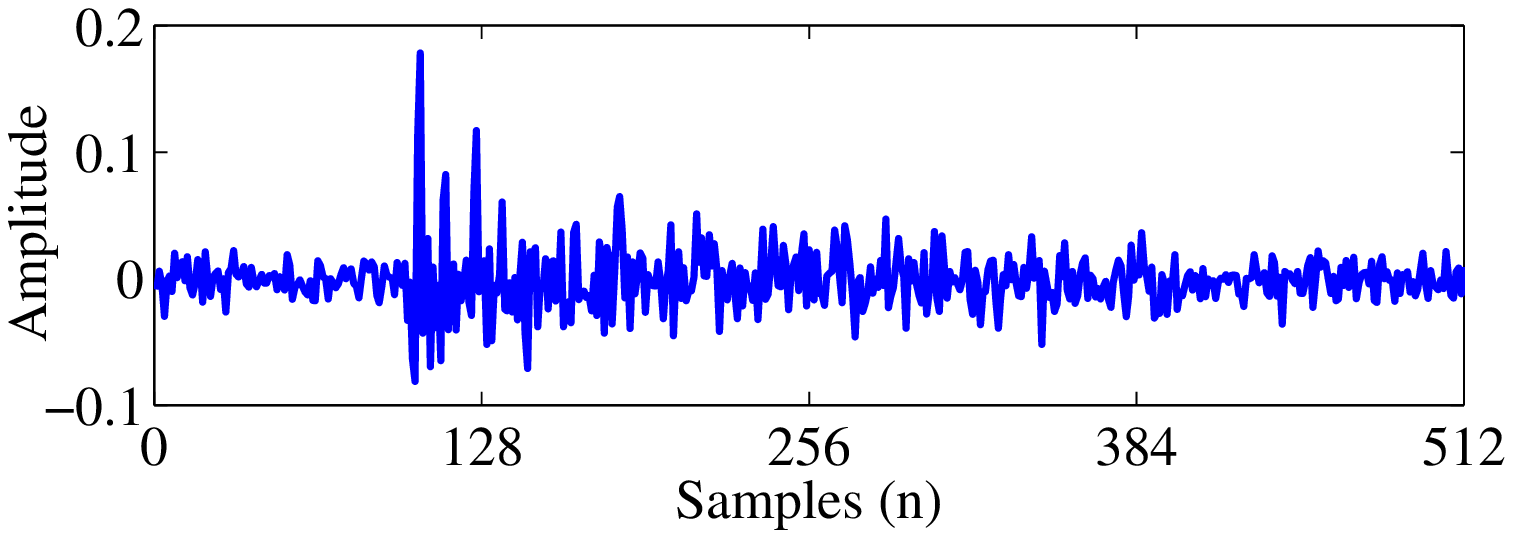} }
\vspace{-2mm} \caption{System impulse responses used in simulations.
} \label{System_Response}
\end{figure*}

\begin{figure*}[t!]
\centering \subfigure[White input sparse system.
\label{non_Sparse}]{
\includegraphics[height=50mm,width=59mm]{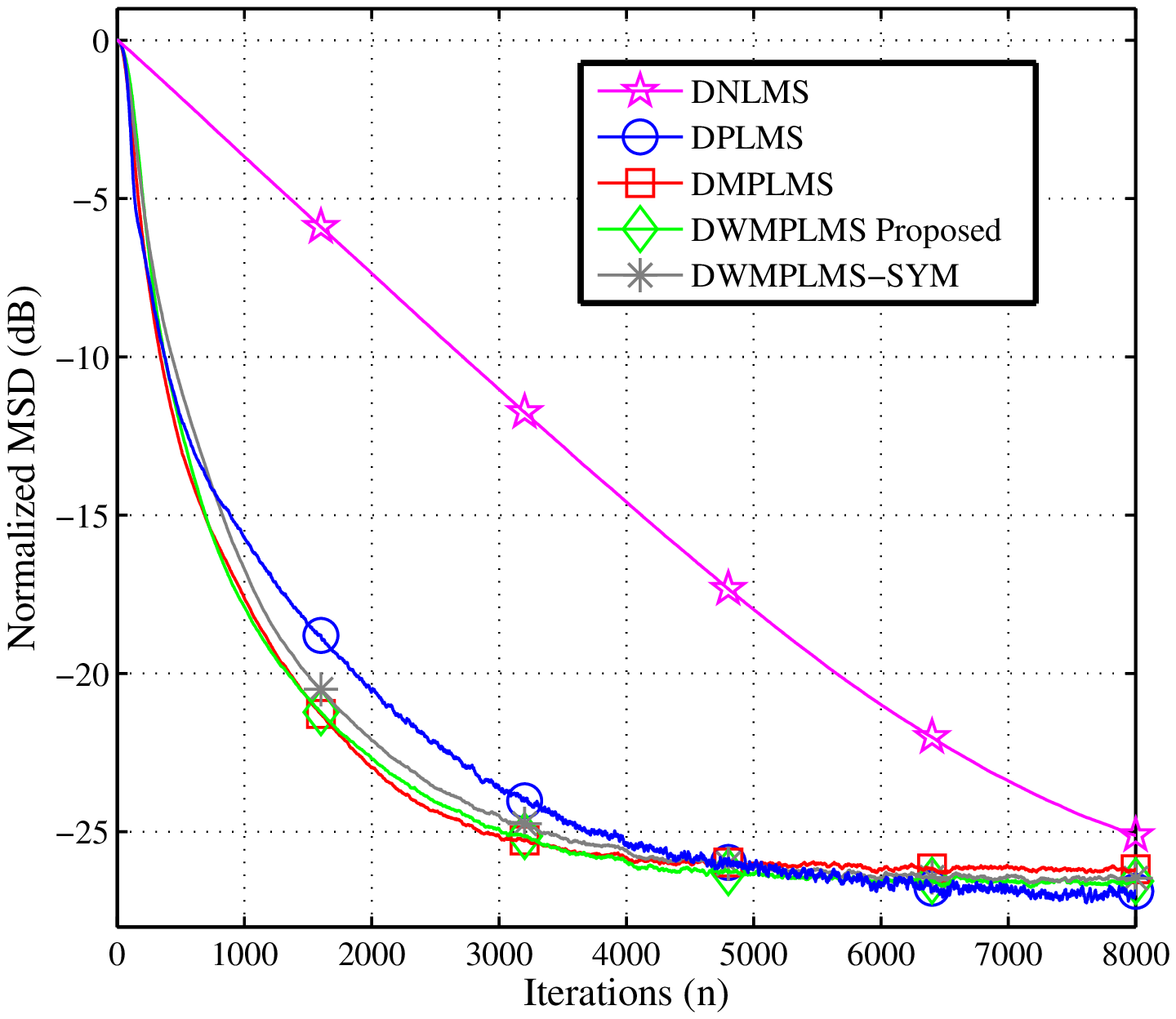} }
\hspace{-5mm} \subfigure[White input semi sparse system.
\label{Semi_Sparse}]{
\includegraphics[height=50mm,width=59mm]{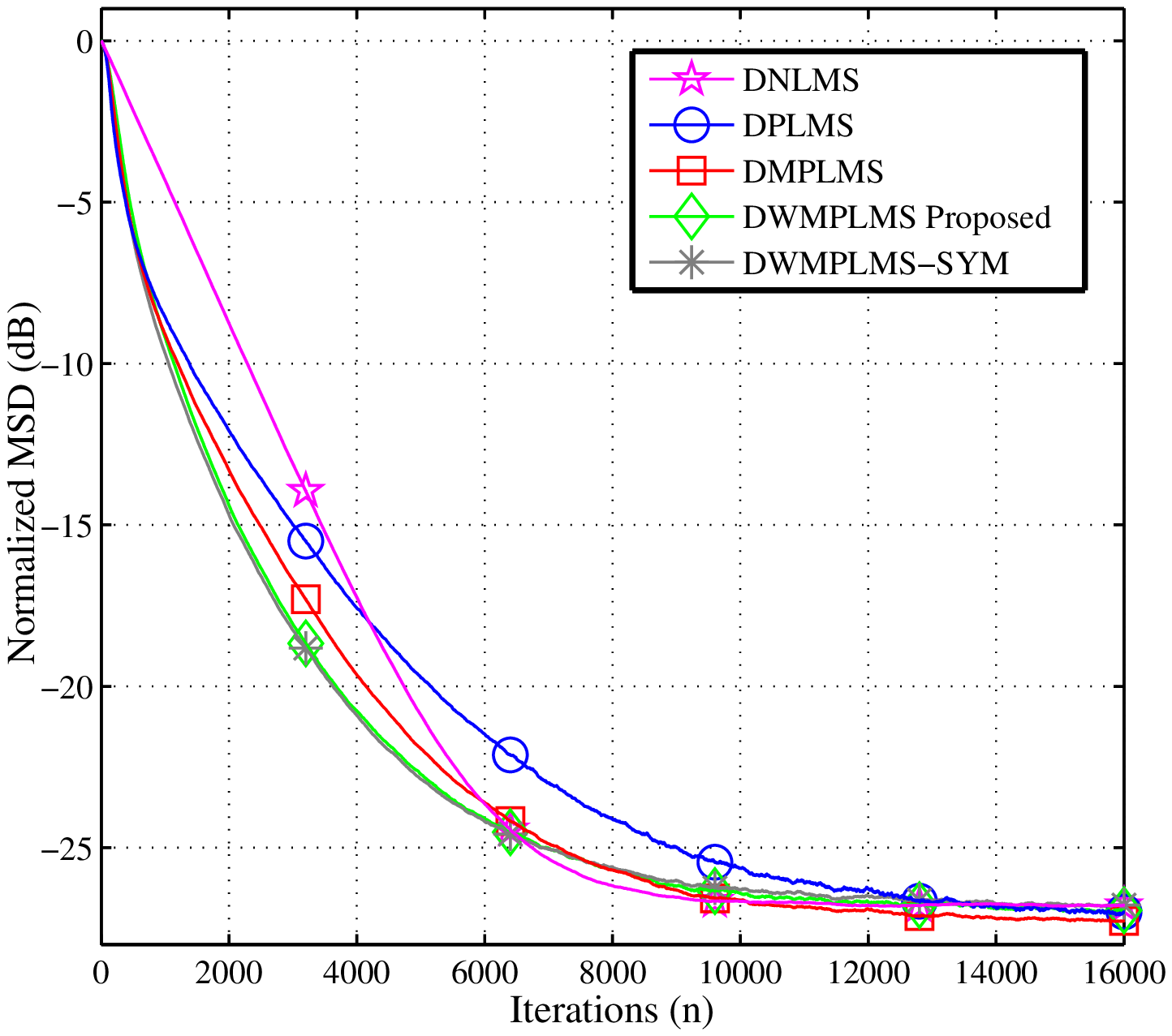} }
\hspace{-5mm} \subfigure[White input disperse system
\label{Sparse}]{
\includegraphics[height=50mm,width=59mm]{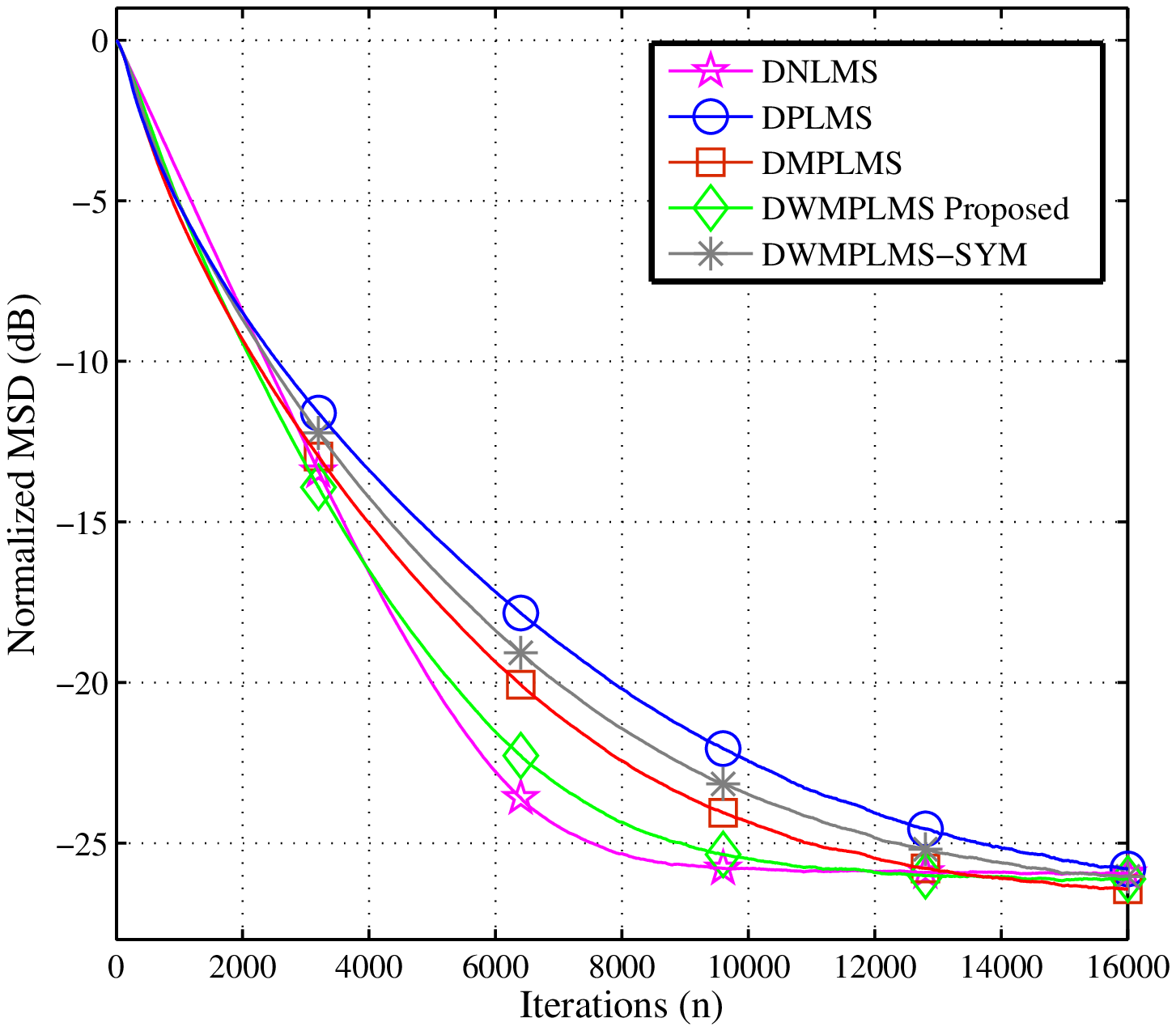} }
\vspace{-2mm} \subfigure[color input sparse system.
\label{Semi_Sparse}]{
\includegraphics[height=50mm,width=59mm]{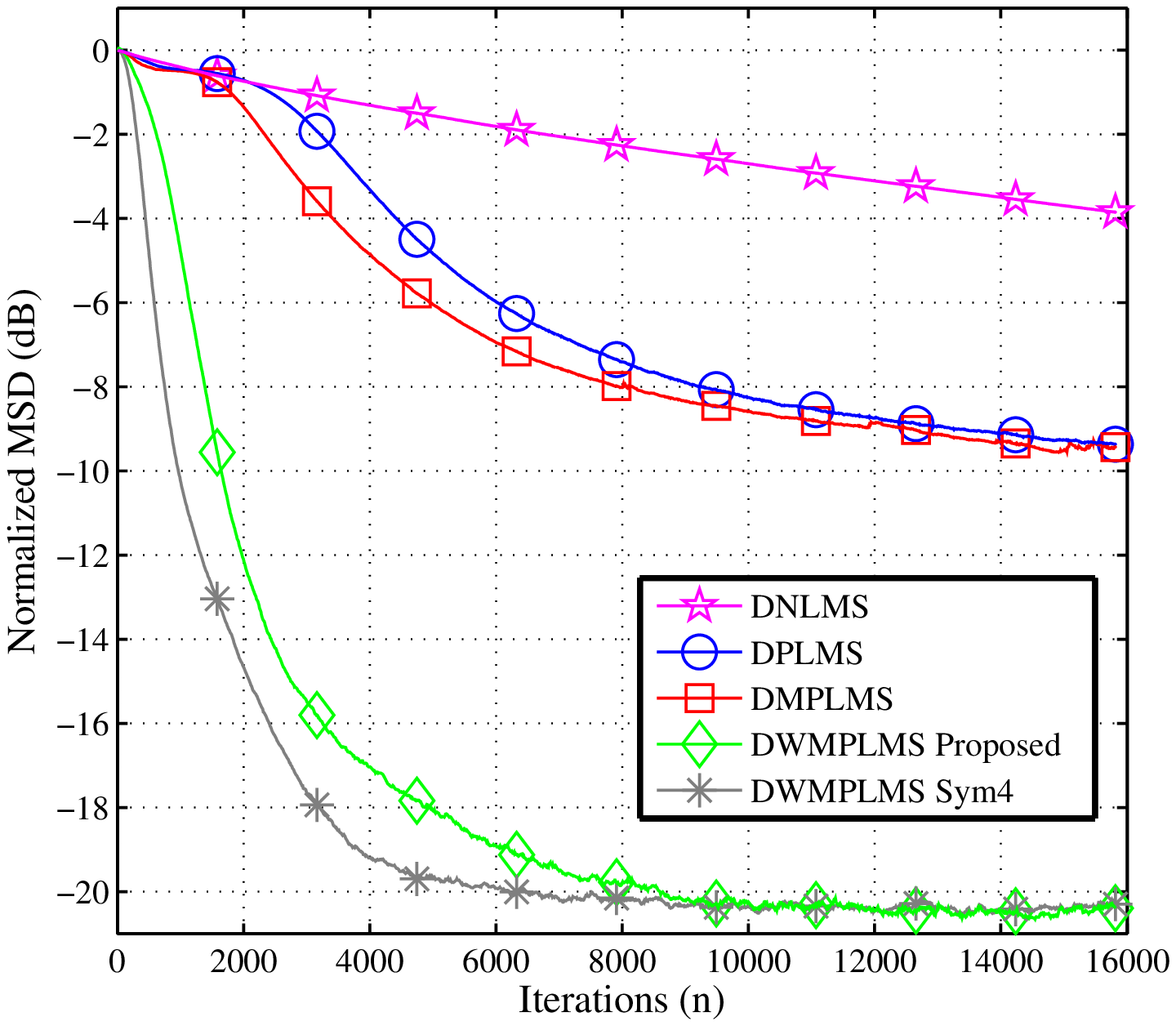} }
\hspace{-5mm} \subfigure[color input semi sparse system.
\label{Semi_Sparse}]{
\includegraphics[height=50mm,width=59mm]{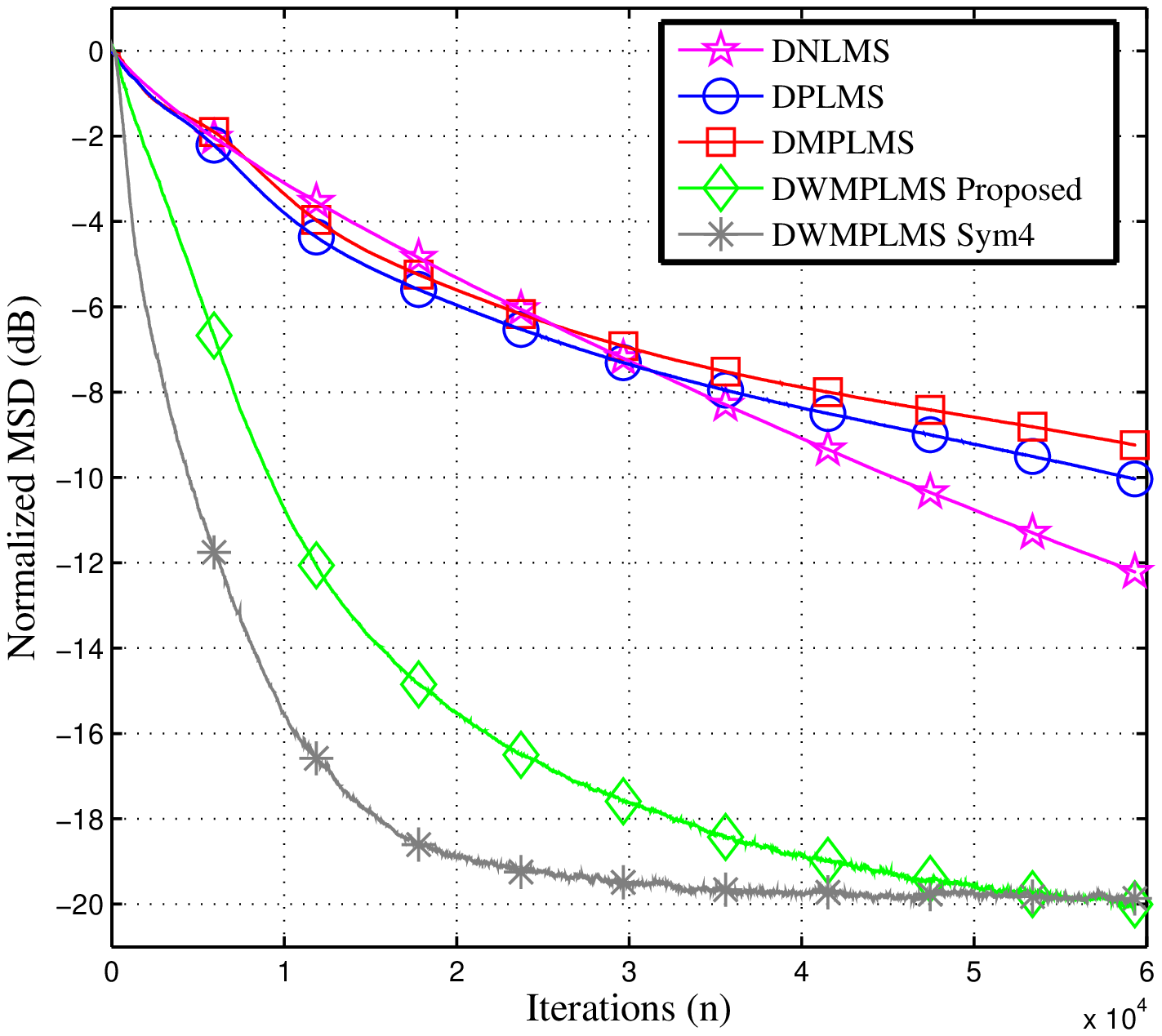} }
\hspace{-5mm} \subfigure[color input disperse system.
\label{Semi_Sparse}]{
\includegraphics[height=50mm,width=59mm]{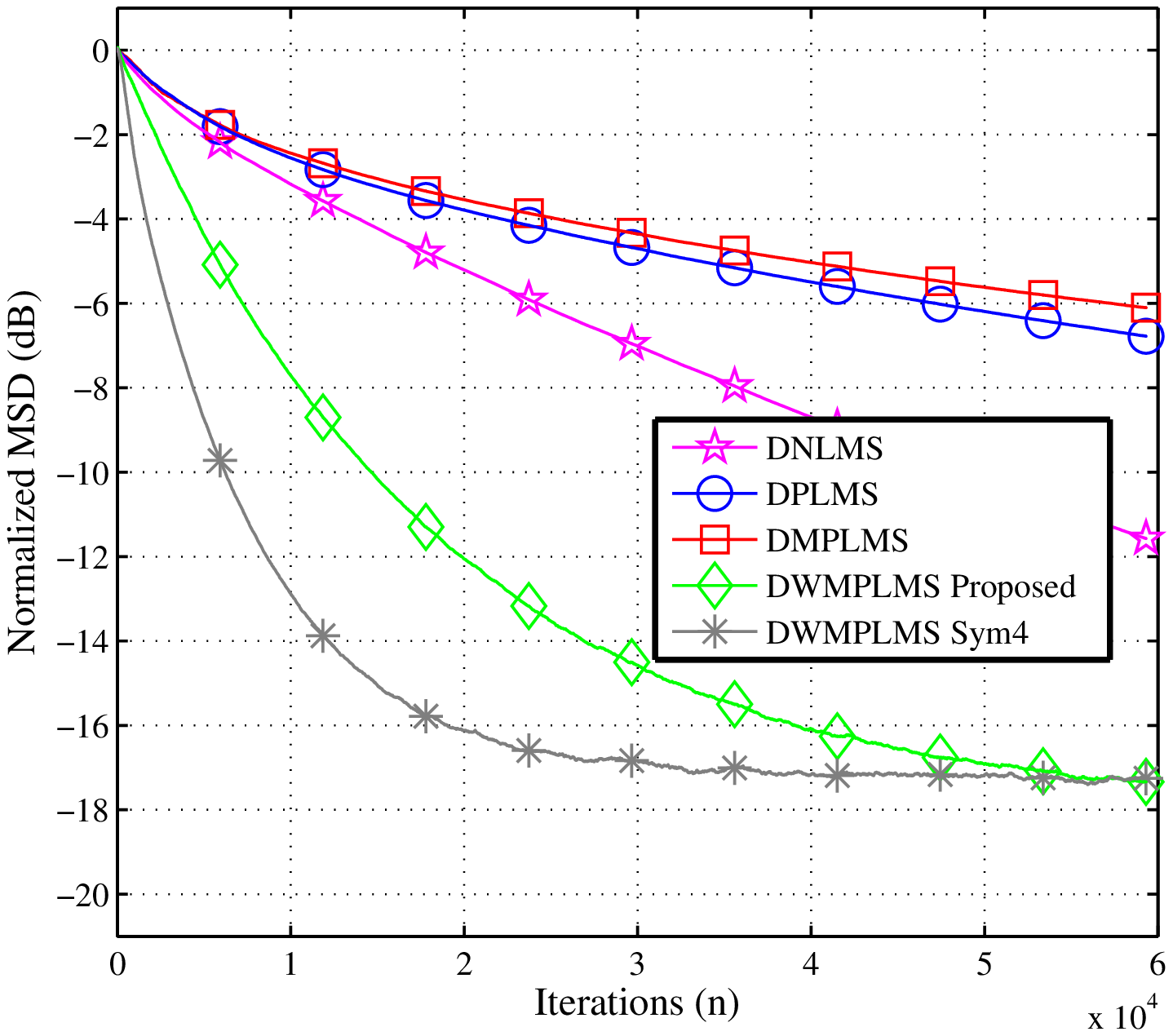} }
\centering \caption{Performance comparisons.
} \label{System_Response sims}
\end{figure*}

\subsection{Absolute power normalization}

In this section we explain the third difference between DMPLMS and DWMPLMS architectures i.e. power normalization.
After applying the orthogonal transform to the input, the input correlation matrix becomes diagonal but it doesn't become identity matrix. Thus we need to normalize the correlation matrix with the power of each tap to make it identity matrix.
The power of each tap is generally estimated using the following equation

\begin{equation}
\Psi_{i}(n) = \beta \Psi_{i}(n-1) + (1-\beta)
 (u_{T,i}(n))^2 , \mspace{4mu} 0\leq i \leq (L-1).
\end{equation}

However, this adds a significant area penalty as it effects all the taps and to simplify this we used absolute of the input instead of the square of the input and the new equation is shown below. We will show that the penalty for this approximation is negligible through simulations which are explained in the next section.
\begin{equation}
\Psi_{i}(n) = \beta \Psi_{i}(n-1) + (1-\beta)
 |(u_{T,i}(n))| , \mspace{2mu} 0\leq i \leq (L-1).
\end{equation}

We choose $\beta = 0.125$ i.e. $\frac{1}{8}$ to avoid multiplication and
this power estimation block is zoomed and shown in Fig.~\ref{dwmpls_arch}.
In the next section we present detailed simulation results for the proposed algorithm.

\begin{figure*}[t!]
\centering \subfigure[Speech i/p.
\label{speech}]{
\includegraphics[height=50mm,width=80mm]{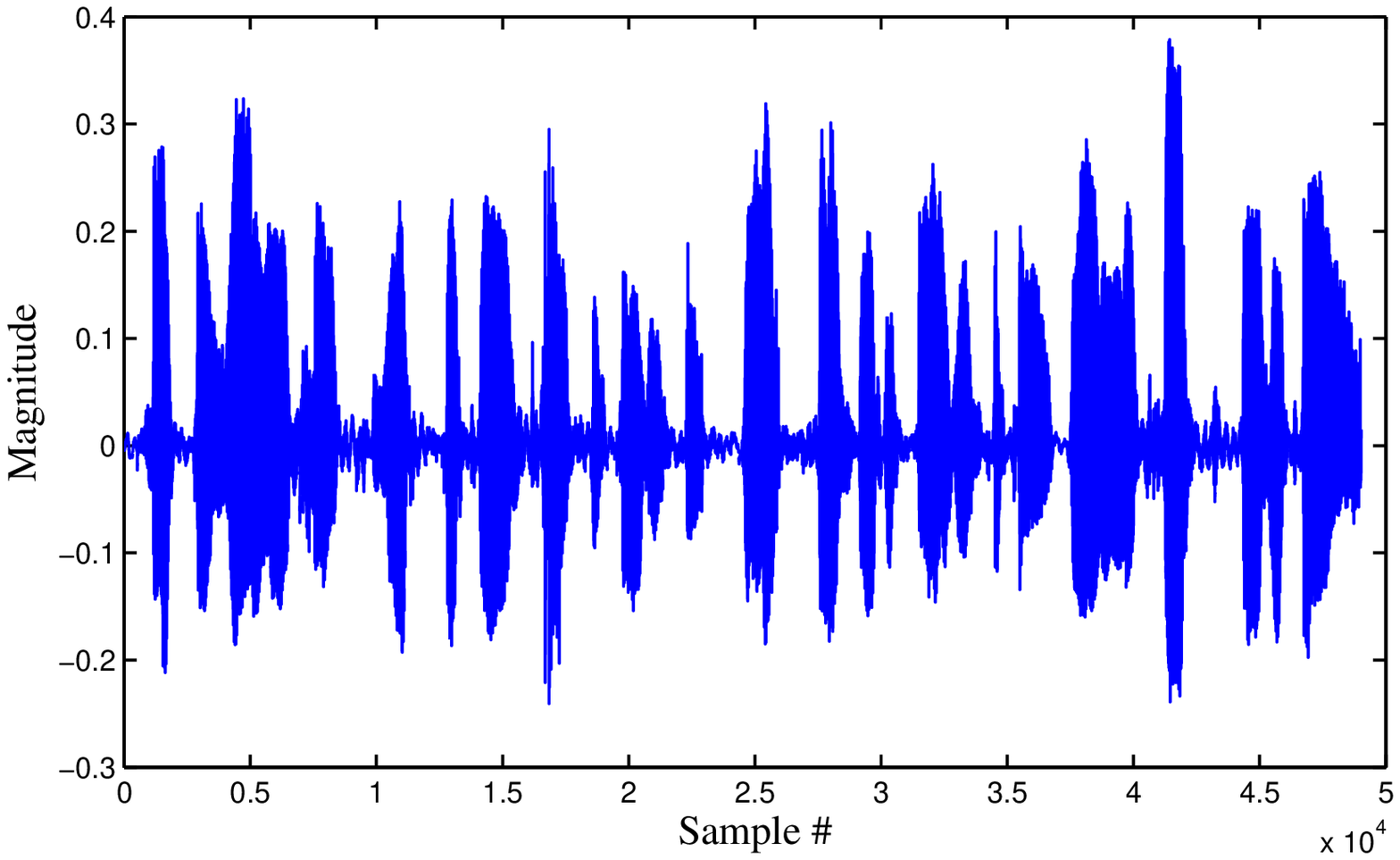} }
\subfigure[Tracking performance.
\label{track_msd}]{
\includegraphics[height=50mm,width=90mm]{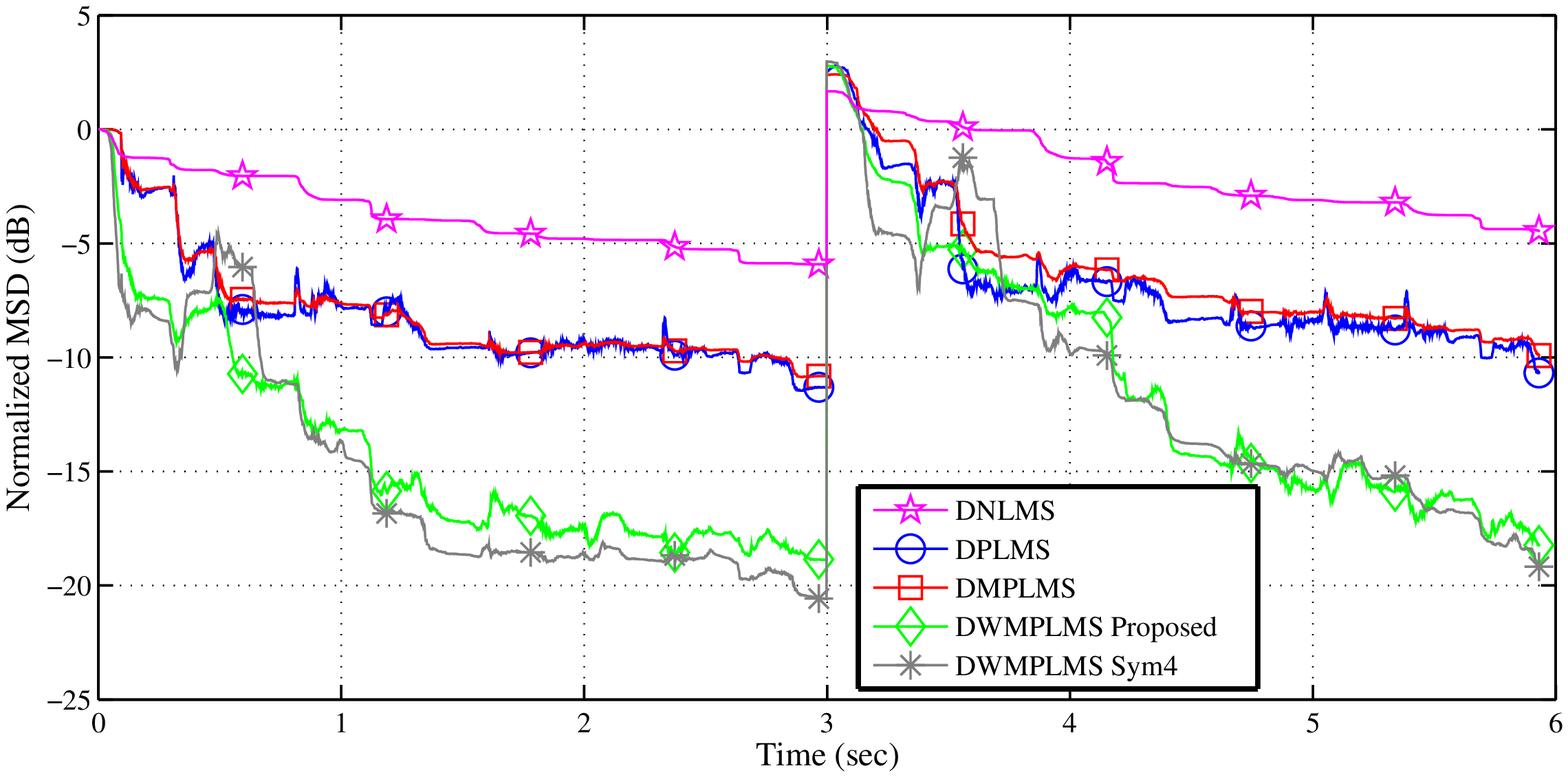} }
 \caption{Performance comparison for speech input.
} \label{Speech_Response}
\end{figure*}

\begin{table*}[t!]
\caption{Synthesis results for TS18 180nm CMOS technology}
\centering
\label{table_synth}
\begin{threeparttable}
\begin{tabular}{ |c|c|c|c|c|c|c|c| }
 \hline

 Algorithm & \# of taps & Clock freq.(MHz) & Leakage Power ($\mu $W) & Dynamic Power (mW) & Cell Area(kGE) \tnote{a} & Core Area($\mu m^2)$   \\

  \hline
  \hline
 \multirow{3}{*}{DLMS}  & $16$ & $100$ & $0.609$ & $38.03$ & $41.1$ & $539290$  \\ \cline{2-7}
 & $32$ & $100$ & $1.21$ & $66.45$ & $82.04$ & $1085032$  \\ \cline{2-7}
 & $64$ & $100$ & $2.43$ & $114.39$ & $164.1$  & $2176258$ \\ \cline{2-7}
  \hline

 \multirow{3}{*}{DPLMS} & $16$ & $100$ & $0.802$ & $33.36$ & $49.57$ & $650031$  \\ \cline{2-7}
 & $32$ & $100$ &  $1.57$ & $61.95$ & $97.26$ & $1280869$   \\ \cline{2-7}
 & $64$ & $100$ & $3.13$ & $119.45$ & $193.47$ & $2558840$  \\ \cline{2-7}
  \hline

 \multirow{3}{*}{DMPLMS} & $16$ & $100$ & $0.867$ & $47.67$ & $52.88$ & $693356$ \\ \cline{2-7}
 & $32$ & $100$ & $1.70$ & $91.72$ & $103.83$ & $1367065$   \\ \cline{2-7}
 & $64$ & $100$ & $3.39$ & $176.97$ & $206.98$ & $2736847$  \\ \cline{2-7}
  \hline

 \multirow{3}{*}{DWMPLMS} & $16$ & $100$ & $1.007$ & $58.31$ & $66.18$ & $863061$ \\ \cline{2-7}
 & $32$ & $100$ & $2.01$ & $115.1656$ & $130.84$ &   $1722799$ \\ \cline{2-7}
 & $64$ & $100$ & $3.39$ & $229.8$ & $260.13$ &   $3441479$\\ \cline{2-7}
  \hline

\end{tabular}
\begin{tablenotes}
       \item[a] One gate equivalent corresponds to the size of a two input NAND gate of size $12.54 \mu m^2$
     \end{tablenotes}
  \end{threeparttable}
\end{table*}

 \section{Simulation Results}

For evaluating the performance of the DWMPLMS algorithm with all the
proposed re-formulations and $3$-level un-normalized Haar wavelet,
Experiments were performed in the context of echo cancelation which
is the main application of sparse adaptive filters. we considered
three systems of varying sparsity. The first system is a network echo path from G168 Recommendation and its impulse response can be considered to be very sparse since the associated sparseness measure
is $0.89$, the second one is a semi sparse system with a sparsity
measure of $0.5$ and third system is a disperse system with a
sparsity of $0.3$. All the impulse responses have $512$
coefficients, using a sampling rate of $8$ kHz. All adaptive filters
used in the experiments have the same length, i.e., L = $512$. The
far-end signal (i.e., the input signal) is either a white Gaussian
signal or a color input obtained by passing a white Gaussian input
through an first order AR process with the pole at $0.95$ and speech
sequence. The output of the echo path is corrupted by an independent
white Gaussian noise (i.e., the background noise at the near-end)
with $30$ dB echo-to-noise ratio (ENR). Adaptation step sizes for color input case are same as those mentioned in \label{3_haar_msd} and for white case, $\mu_{dnlms} = \mu_{dmplms} = \mu_{dwmplms-sym4} = 0.25 , \mu_{dplms} = 0.22, \mu_{dwmplms-proposed} = 0.15$. Step sizes are adjusted such that steady state MSD is equal. We can see from Fig.~\ref{System_Response sims}, For both color and white inputs, as the sparsity decreases performance of DPLMS/DMPLMS becomes worse than that of DNLMS where as DWMPLMS is able to perform better or equal to the DNLMS. For the network echo cancelation which comes under color input sparse system, DWMPLMS significantly outperforms all the other three. We see that even if the sparsity or input correlation varies over wide range, DWMPLMS is able to perform consistently. We also evaluated the tracking performance (by changing the echo path in the midway by shifting the impulse response to the right by $12$ samples) of the proposed algorithm using the speech input shown in Fig.~\ref{speech} for the sparse system and result is shown in Fig.~\ref{track_msd}. All other simulation parameters are same as that of color case and $\mu_{dnlms} = 0.2, \mu_{dmplms} =\mu_{dplms} = 50, \mu_{dwmplms-sym4} = 2.5, \mu_{dwmplms-proposed} = 1.1$. Please note that we need to adjust the step sizes depending on the input variance which is very low for the speech input and hence large value of $\mu$ for DMPLMS and DPLMS. We can see that the performances are similar to color case and we can conclude that DWMPLMS with U-HAAR 3-level decomposition is the robust VLSI solution to the echo cancelation with varying sparsity.

%
%

 \section{VLSI Implementation results and comparative study}

In this section we provide the synthesis results of
the proposed DWMPLMS, DMPLMS and DPLMS architectures and since there are no previous architectures reported for them in literature, we compare the results  with standard DLMS to see how much is the complexity increase for the given convergence performance improvement. All the designs are implemented in Verilog, synthesized using Synopsys Design Compiler with TS$18$
standard cell library (Tower Semiconductor $180$nm CMOS technology).
$16$-bit fixed point representation is considered for all the
designs. Filter lengths of $16, 32$ and $64$ are considered for comparing the scalability of time
and area complexities and all the designs are targeted for $100$ MHz
for fair comparison. The results are summarized in Table.~\ref{table_synth}.

Compared to standard DLMS, re-formulated DPLMS has $20.53\%, 18.04\%$ and $17.57\%$ more area complexity for $16, 32$ and $64$ taps respectively. We see that the complexity increase is not exactly linear and it is intuitive because of overheads like adder tree, error calculation and G calculation blocks. Similarly DMPLMS has $28.56 \%, 25.99 \%, 25.75 \%$ area complexity increase compared to DLMS. So roughly with $25\%$ increase in hardware we are getting 3x improvement in convergence performance in case of white input. Finally for color input when we consider DWMPLMS, it has $60.03 \%,58.77 \%, 58.13 \%$ area increase compared to DLMS complexity for $16, 32$ and $64$ taps respectively. So we can conclude that with $58 \%$ increase in hardware we are able to achieve 15x improvement in convergence for color input.

Finally we realized one higher order filter with 256-tap for the proposed DWMPLMS and the synthesis results at $100$MHz target frequency are shown in Table.~\ref{high_ord} and the detailed area break-up is shown in Table.~\ref{high_area}. We see that filter taps occupy most of the of area, followed by sliding HAAR unit which occupy around $10\%$ of the area. However, one can notice that the core area is on the higher side but thats because of $180$nm process and with more advanced technology nodes it will be scaled down accordingly.

\begin{table}[h]
\caption{Synthesis Results of $256$-tap DWMPLMS}
\centering
\begin{tabular}{ll}
\Xhline{2\arrayrulewidth}

 Parameter  & Value  \\
\Xhline{2\arrayrulewidth}
Clock freq    & $100$ MHz   \\ [0.5ex]
Dynamic Power          &  $417$ mW           \\
Leakage Power       &   $16.9$ $\mu$W      \\
Cell Area      &    $1037$  kGE  \\
Core Area  &  $13.78$ mm$^2$   \\
\Xhline{2\arrayrulewidth}
\end{tabular}
\label{high_ord}
\end{table}

\begin{table}[h]
\caption{Area Break Down of the Sub-blocks}
\centering
\begin{tabular}{lll}
\Xhline{2\arrayrulewidth}
  & kGE & \% \\
\hline
$\text{E Calculation}$  & $1.62$ & $0.16$  \\
Adder Tree      &  $27.98$ & $2.7$           \\
$256$ Taps   &    $898.75$ & $86.62$  \\
Sliding HAAR unit      &  $109.1$ & $10.52$           \\
\hline
Total  &    $1037.45$ & $100$  \\
\Xhline{2\arrayrulewidth}
\end{tabular}
\label{high_area}
\end{table}


 \section{Conclusions and Future Work}

Proportionate-type adaptive algorithms can significantly improve the
convergence performance of the sparse adaptive filters compared to
conventional LMS algorithm. However, the huge computational penalty
associated with these algorithms make their VLSI realization a
highly challenging task even with the most advanced technology
nodes. By utilizing the fact that stochastic gradient algorithms
are tolerant to approximations, we proposed several re-formulations
in this paper to simplify the original PNLMS algorithms and compared
the performances of these algorithms with those of the original
algorithm and proposed an efficient VLSI architectures for these
re-formulated algorithms. We also demonstrated that the DWMPLMS with the proposed $3$-level un-normalized HAAR wavelet is a robust VLSI solution
for the practical echo cancelers with time varying sparsity. This is the first attempt to implement proportionate type algorithms in hardware and this result will
 motivate other researchers to explore more efficient hardware solutions to further improve the sparse adaptive filter architectures.

\vfill

\end{document}